\documentclass[useAMS,usenatbib]{mn2e}
\usepackage{graphicx,amssymb}
\newcommand{\bc}{\begin{center}}
\newcommand{\ec}{\end{center}}
\newcommand{\tdyn}{\tau_{\rm dyn}}
\newcommand{\rvir}{R_{\rm vir}}
\newcommand{\mvir}{M_{\rm vir}}
\newcommand{\vvir}{V_{\rm vir}}

\title[A SAM comparison - gas cooling and galaxy mergers]
      {A semi-analytic model comparison - gas cooling and galaxy mergers}
\author[G.~De Lucia et al.]
       {Gabriella De Lucia$^{1}$\thanks{Email: delucia@oats.inaf.it}, 
        Michael Boylan-Kolchin$^{2}$, Andrew J. Benson$^{3}$, 
        Fabio \newauthor Fontanot$^{1}$,  Pierluigi Monaco$^{1,4}$
        \\  
        $^1$INAF - Astronomical Observatory of Trieste, via G.B. Tiepolo 11, 
        I-34143 Trieste, Italy\\
        $^2$Max--Planck--Institut f\"ur Astrophysik, 
        Karl--Schwarzschild--Str. 1, D-85748 Garching, Germany\\
        $^3$MC350-17, California Institute of Technology, Pasadena, CA 91125,
        USA\\ 
        $^4$Dipartimento di Astronomia, Universit\`a di Trieste, via
        G.B. Tiepolo 11, I-34131 Trieste,Italy} 
\begin{document}

\pagerange{\pageref{firstpage}--\pageref{lastpage}} 
\pubyear{2010}

\maketitle

\label{firstpage}

\begin{abstract}
  We use stripped-down versions of three semi-analytic galaxy formation models
  to study the influence of different assumptions about gas cooling and galaxy
  mergers.  By running the three models on identical sets of merger trees
  extracted from high-resolution cosmological $N$-body simulations, we are able
  to perform both statistical analyses and halo-by-halo comparisons. Our study
  demonstrates that there is a good statistical agreement between the three
  models used here, when operating on the same merger trees, reflecting a
  general agreement in the underlying framework for semi-analytic models. We
  also show, however, that various assumptions that are commonly adopted to
  treat gas cooling and galaxy mergers can lead to significantly different
  results, at least in some regimes. In particular, we find that the different
  models adopted for gas cooling lead to similar results for mass scales
  comparable to that of our own Galaxy. Significant differences, however, arise
  at larger mass scales. These are largely (but not entirely) due to different
  treatments of the `rapid cooling' regime, and different assumptions about the
  hot gas distribution. At this mass regime, the predicted cooling rates can
  differ up to about one order of magnitude, with important implications on the
  relative weight that these models give to AGN feedback in order to
  counter-act excessive gas condensation in relatively massive haloes at low
  redshift. Different assumptions in the modelling of galaxy mergers can also
  result in significant differences in the timings of mergers, with important
  consequences for the formation and evolution of massive galaxies.  
\end{abstract}


\section{Introduction}
\label{sec:intro}

Understanding how galaxies form and the physics that drive their evolution has
been a long-standing problem in modern astrophysics. A number of observational
tests have recently succeeded in determining the fundamental cosmological
parameters with uncertainties of only a few per cent, thus effectively removing
a large part of the parameter space in galaxy formation studies. We are left,
however, with the problem of dealing with our `ignorance' of complex physical
processes, that are inter-twined in an entangled network of actions,
back-reactions, and self-regulations.

Over the past decades, different methods have been developed to study galaxy
formation in a cosmological context. Among these, semi-analytic models have
turned into a flexible and widely used tool to provide detailed predictions of
galaxy properties at different cosmic epochs. These techniques find their seeds
in the pioneering work by \citet{White_and_Rees_1978}, were laid out in a more
detailed form in the early 1990s
\citep*{White_and_Frenk_1991,Cole_1991,Kauffmann_White_Guiderdoni_1993}, and
have been substantially extended and refined in the last years by a number of
different groups \citep[for a review, see][]{Baugh_2006}. In these models, the
evolution of the baryonic components is modelled adopting simple yet physical
and/or observationally motivated {\it recipes}, coupled in a set of
differential equations that describe the variation in mass as a function of
time of different galactic components (e.g. stars, gas, metals).

While it is relatively easy to compare results from different models\footnote{A
  number of galaxy catalogues have been made publicly available by various
  groups; results from different versions of two of the models used in this
  study are available through a relational database accessible at:
  \\ http://www.mpa-garching.mpg.de/millennium/}, it is more complicated to
understand the origin of any difference or similarity between them. This
difficulty stems primarily from the fact that different groups adopt different
sets of prescriptions (that are equally reasonable, given our poor
understanding of the physics at play) and that, as mentioned above, the final
results are given by a combination of these. There are, however, also a number
of more subtle differences that are more `technical' in nature (e.g. the
particular mass definition adopted, the cooling functions used, the use of
analytic or numerical merger trees, etc.). The precise influence on models'
results of at least some of these details is unclear. For example, it has been
shown that the extended Press-Schechter (EPS) formalism
\citep{Bond_etal_1991,Bower_1991} does not provide an adequate description of
the merger trees extracted directly from numerical simulations
\citep{Benson_Kamionkowski_Hassani_2005,Cole_etal_2008}. Although some of these
most recent studies have provided `corrections' to analytic merger trees, many
applications are still carried out using the classical EPS formalism, and
little work has been done to understand to which measure this affects
predictions of galaxy formation models.

In this paper, we compare results from three independently developed
semi-analytic models. Our goal is not to predict or reproduce any specific
observation. Rather our aim is to understand the level of agreement between
different semi-analytic models, with a minimum of assumptions or free
parameters. This requires (1) that the models are implemented on {\it
  identical} merger trees, such that results can be compared on a case-by-case
basis and any differences can be attributed to {\it specific parametrizations
  of physical processes}; and (2) that a minimum of physical processes are
included in the models in order to avoid possible degeneracies, and to hopefully
illuminate the effects of specific parametrizations or parameter choices.

In our study, the first requirement has been satisfied by creating a standard
set of halo merger trees extracted from $N$-body simulations, and running each
model on these trees. The second requirement is somewhat more demanding as
modern semi-analytic models contain treatment of numerous, coupled physical
processes. We have chosen to simplify the models as much as possible by
removing {\it all} physics other than gas cooling and galaxy mergers. This
allows us to focus on the influence of different assumptions typically made to
model these two physical processes, that represent two basic ingredients of any
galaxy formation model. Using large samples of identical haloes, we are able to
compare results both in a statistical fashion and on a halo-by-halo basis.

Previous studies \citep[e.g.][]{Benson_et_al_2001, Yoshida_et_al_2002,
  Helly_etal_2003, Cattaneo_et_al_2007} have compared numerical predictions
from stripped-down versions of semi-analytic models with those from
hydrodynamical simulations, to verify whether these methods provide consistent
predictions in the idealized case in which only gas cooling is included. The
general consensus is that the cooling model usually employed in semi-analytic
models is in good agreement with hydrodynamical simulations that adopt the same
physics. More recent studies focused on object-by-object comparisons, however,
have highlighted a number of important differences that were `hidden' in the
relatively good agreement obtained by previous studies focusing on statistical
comparisons \citep{Saro_etal_2010}. A recent study by \citet{Viola_etal_2008},
in particular, has shown that the cooling model implemented in {\small MORGANA}
(one of the models used in this study) predicts cooling rates that are
significantly larger than those predicted from their implementation of the
`classical' cooling model, proposed by \citet{White_and_Frenk_1991}. In
addition, Viola et al. have shown that {\small MORGANA} provides results that
are in good agreement with those of controlled numerical experiments of
isolated haloes, with hot gas in hydrostatic equilibrium. While both SPH and
semi-analytic techniques have their own weaknesses, making it unclear which of
the two (if either) is providing the `correct' answer, these results appear
confusing. It is therefore interesting to study how the different possible
assumptions that can be made to model the evolution of cooling gas, propagate
on predictions from galaxy formation models.

Modelling of galaxy mergers has not been considered a major concern, but
different assumptions about merging time-scales can be made, and these may have
important consequences for the inferred stellar assembly history of massive
galaxies, including brightest cluster galaxies. In addition, recent work has
shown that the classical dynamical friction formula usually adopted in
semi-analytic models tends to under-estimate merging times computed from
controlled numerical experiments and high-resolution cosmological simulations
(\citealt*{Boylan-Kolchin_Ma_Quataert_2008}, \citealt{Jiang_etal_2008}).
Results from these studies, however, have not yet converged on the appropriate
correction(s) that should be applied to the classical formula.

In this paper, we will not address the issue of what is {\it the best way} to
model galaxy mergers or gas cooling.  Instead, we will concentrate on the
differences due to alternative implementations of these physical processes,
with the aim of understanding their effects in a full semi-analytic model.  We
will also explore what these effects might imply for the importance of other
physical processes.

The numerical simulations and merger trees used in our study are described in
Section \ref{sec:simstrees}. In Section \ref{sec:models}, we describe in detail
how gas cooling and galaxy mergers are treated in each of the three models used
in this work. Section \ref{sec:mw} and Section \ref{sec:scuba} present our
results for two halo samples. In Section \ref{sec:ressub}, we discuss the
influence of numerical resolution and of different schemes for the construction
of merger trees. Section \ref{sec:mergtimes} compares the different
implementations of merger times adopted in the three models
considered. Finally, we summarize and discuss our results, and give our
conclusions in Section \ref{sec:disc}.

\section{The simulations and the merger trees}
\label{sec:simstrees}

This work takes advantage of two large high-resolution cosmological
simulations: the {\it Millennium Simulation} (MS,
\citealt{Springel_etal_2005}), and the {\it Millennium-II} (MS-II,
\citealt{Boylan-Kolchin_etal_2009}). The MS follows $N=2160^3$ particles of
mass $8.6\times10^{8}\,h^{-1}{\rm M}_{\odot}$, within a comoving box of size
$500\, h^{-1}$Mpc on a side. The MS-II follows the evolution of the same number
of particles in a volume that is 125 times smaller than for the MS ($100\,
h^{-1}$Mpc on a side). The particle mass is correspondingly 125 times smaller
than for the MS ($6.9\times10^{6}\,h^{-1}{\rm M}_{\odot}$), allowing haloes
similar to those hosting our Milky Way to be resolved with hundreds of
thousands of particles. For both simulations, the cosmological model is a
$\Lambda$CDM model with $\Omega_{\rm m}=0.25$, $\Omega_{\rm b}=0.045$,
$h=0.73$, $\Omega_\Lambda=0.75$, $n=1$, and $\sigma_8=0.9$. The Hubble constant
is parameterised as $H_0 = 100\, h\, {\rm km\, s^{-1} Mpc^{-1}}$. In order to
test how numerical resolution affects our results, we will also use the {\it
  mini-Millennium-II} simulation \citep[mini-MSII; ][]{Boylan-Kolchin_etal_2009},
which was run using the same initial conditions and volume as for the MS-II,
but at the mass and force resolution as for the original MS (the number of
particles is therefore $432^3$). The basic properties of the three simulations
used in this study are summarised in Table~\ref{tab:sims}.

\begin{table}
\begin{tabular}{lccc}
 \hline
 {\bf Name}
 & $L_{\rm box}$
 & $\epsilon$ [$h^{-1}\,{\rm kpc}$]
 & $m_p$ [$h^{-1}{\rm M}_{\odot}$]\\
 \hline
 MS & 500 & 5.0  & $8.61\times 10^{8}$\\
 MS-II & 100 & 1.0 & $6.89 \times 10^{6}$\\
 mini-MS-II & 100 & 5.0 &  $8.61\times 10^{8}$\\
\hline
\end{tabular}
\caption{ Some basic properties of the three simulations used in this study:
  the side length of the simulation box $L_{\rm box}$, the Plummer-equivalent
  force softening $\epsilon$, and the particle mass $m_p$.}
\label{tab:sims}
\end{table}

For each simulation snapshot (64 for the MS, 68 for the MS-II), group
catalogues were constructed using a standard friends--of--friends (FOF)
algorithm, with a linking length of $0.2$ in units of the mean particle
separation. Each group was then decomposed into a set of disjoint substructures
using the algorithm {\small SUBFIND} \citep{Springel_etal_2001}, which
iteratively determines the self-bound subunits within a FOF group.  The most
massive of these substructures is often referred to as the {\it main halo},
while this and all other substructures are all referred to as {\it subhaloes}
or {\it substructures}. Only subhaloes that retain at least $20$ bound
particles after a gravitational unbinding procedure are considered `genuine'
subhaloes, and are used to construct merger history trees as explained in
detail in \citet{Springel_etal_2005} (see also
\citealt{DeLucia_and_Blaizot_2007} and \citealt{Boylan-Kolchin_etal_2009}). The
subhalo detection limit is therefore set to $2.36\times10^{10}$, and to
$1.89\times10^8\,{\rm M}_{\odot}$ for the MS and the MS-II, respectively.  Note
that some FOF haloes do not contain 20 self-bound particles; such FOF haloes
are not included in the merger trees.

The comparison discussed below will focus on two samples of haloes. The first
sample consists of 100 haloes, selected from the MS-II on the basis of their
mass at $z=0$, with $\log {\rm M}_{200}$ between $11.5$ and $12.5$. Here ${\rm
  M}_{200}$ is in units of $h^{-1}{\rm M}_{\odot}$, and is defined as the mass
within a sphere of density 200 times the critical density of the Universe at
the corresponding redshift\footnote{The corresponding radius, $R_{200}$,
    has been shown to approximately demarcate the inner regions of haloes which
    are in dynamical equilibrium, from the outer regions which are still
    infalling \citep{Cole_and_Lacey_1996}. In the following, we will refer to
    this radius as the `virial radius', $\rvir$. The virial mass $\mvir$ is the
    mass contained within the sphere defined by this radius, and the virial
    velocity $\vvir$ is the circular velocity at $\rvir$.}. We will refer to
this sample as the `Milky-Way like' sample. A second sample of other 100 haloes
was selected from the MS by taking haloes that have a number density of
$10^{-5}$ at $z\sim 2$, and that end up in massive groups/clusters at
$z=0$. The adopted number density is comparable to that of submillimiter
galaxies at $z\sim 2$ \citep[][and references therein]{Chapman_etal_2004}. We
will refer to this sample as the `{\small SCUBA}-like' sample.

As mentioned above, trees for the MS and MS-II were originally constructed at
the subhalo level.  Many semi-analytic models, however, are based on FOF merger
trees.  We have therefore constructed FOF-based merger trees for our halo
samples.  There are many possible ways to construct FOF merger trees from
subhalo trees.  We have chosen one of the most straightforward methods: since
there exists a one-to-one correspondence between FOF haloes in the subhalo
trees and main subhaloes, we are free to assign the merger history of a FOF
halo to the merger history of its dominant subhalo.  We use this mapping from
dominant subhaloes to FOF haloes to directly link the merger history of
dominant subhaloes to the merger history of their host FOF haloes.  So, for
example, if dominant subhalo $\alpha$ of FOF group $A$ has the descendant
$\beta$ in FOF group B, then FOF halo $B$ is defined to be the descendant of
FOF halo $A$ in our FOF trees.

More sophisticated algorithms for constructing FOF trees from the MS or MS-II
subhalo trees exist \citep[e.g.][]{Fakhouri_and_Ma_2008,Genel_etal_2008}.  In
particular, these algorithms filter out ``unphysical'' FOF mergers -- for
example, those due to chance associations of two FOF haloes for only one
snapshot.  Each of these algorithms has different strengths and weaknesses. 
The primary virtue of the algorithm we have chosen is its simplicity, and the
advantage that each FOF tree can be easily connected to the corresponding
subhalo tree (we will use subhalo-based merger trees in Section
\ref{sec:ressub}).

\section{The models}
\label{sec:models}

In this study, we use three different and independently developed semi-analytic
codes. In the following, we will refer to them as the {\it Munich} model, the
{\it Durham} model, and \emph{\small MORGANA}. As explained above, we have used
stripped-down versions of the models that only include gas cooling and galaxy
mergers, so as to focus on a few specific aspects of the modelling. In
addition, all models have been adapted to run on the merger trees described in
Section \ref{sec:simstrees}. In the following, we describe in more details how
gas cooling and galaxy mergers are treated in each of the models used in our
study, and the changes that were made in order to adapt each code to the same
merger trees.

\subsection{The Munich model}

The version of the Munich model used in this study is the one described in
\citet{DeLucia_and_Blaizot_2007}, and we refer to the original paper and
references therein for more details.

The rate of gas cooling is computed following the model originally proposed
\citet{White_and_Frenk_1991}, and an implementation similar to that adopted in
\citet{Springel_etal_2001}. The Munich model assumes that the hot gas within
dark matter haloes follows an isothermal profile:
\begin{displaymath}
\rho_{\rm g}(r) = \frac{M_{\rm hot}}{4\pi R_{200} r^2}.
\end{displaymath}

For each new snapshot, the total amount of hot gas available for cooling in
each halo is estimated as follows:
\begin{equation}
  M_{\rm hot} = f_{\rm b}\mvir-\sum_i M_{\rm cold}^{(i)}
\label{eq:mhotcompute}
\end{equation}
where the sum extends over all the galaxies in the FOF halo and $f_{\rm b}$ is
the baryon fraction of the Universe, for which we assume the value
0.017\footnote{In this work, all models assume the same value for the universal
  baryon fraction.}. Eq.~\ref{eq:mhotcompute} can provide, in a few cases, a
negative number (this occurs typically after important halo mergers). In this
case, the amount of hot gas is set to zero, and no cooling is allowed in the
remnant halo.

The equations driving galaxy evolution are then solved using $20$ time-steps
between each pair of simulation snapshots. A local cooling time is defined as
the ratio between the specific thermal energy content of the gas and the
cooling rate per unit volume:
\begin{equation}
  t_{\rm cool}(r) = \frac{3}{2} \frac{kT\rho_{\rm g}(r)}{\bar{\mu}m_{\rm
  p}n_{\rm e}^2(r)\Lambda(T,Z)}
\label{eq:tcool}
\end{equation}
In the above equation, $\bar{\mu}m_{\rm p}$ is the mean particle mass, $n_{\rm
  e}(r)$ is the electron density, $k$ is the Boltzmann constant, and
$\Lambda(T,Z)$ represents the cooling rate. The latter is strongly dependent on
the virial temperature of the halo, and on the metallicity of the gas. In the
Munich model, these dependencies are accounted for by using the collisional
ionization cooling curves by \citet{Sutherland_and_Dopita_1993}. Since chemical
enrichment is switched off in this study, we only use the calculation
corresponding to `primordial' composition. The virial temperature of the halo
is determined using the hydrostatic equilibrium equation, and relates the gas
temperature to the circular velocity of the halo:
\begin{displaymath}
T_{\rm vir} = \frac{1}{2} \frac{\mu m_{\rm H}}{k} \vvir^2\,\,\,\,\,\,\,{\rm
  or}\,\,\, T_{\rm vir} = 35.9(\vvir/{\rm km}\,{\rm
  s}^{-1})^2\,\,\,{\rm K}
\end{displaymath}
where $m_{\rm H}$ is the mass of the hydrogen atom, and $\mu$ is the mean
molecular mass.

A `cooling radius' is then computed as the radius at which the local cooling
time is equal to the halo dynamical time. We note that in the original work by
White \& Frenk, the cooling radius was defined equating the local cooling time
to the age of the Universe, which is about one order of magnitude larger than
the halo dynamical time. As discussed in \citet*{DeLucia_Kauffmann_White_2004},
the particular choice currently adopted in the Munich model was motivated by
the significant enhancement of cooling rates when adopting metal dependent
cooling functions.

If the cooling radius lies within the virial radius of the halo under
consideration, the gas is assumed to cool quasi-statically, and the cooling
rate is modelled by a simple inflow equation:
\begin{displaymath}
\frac{{\rm d}M_{\rm cool}}{{\rm d}t} = 4\pi\rho_{\rm g}(r_{\rm cool})r_{\rm
  cool}^2\frac{{\rm d}r_{\rm cool}}{{\rm d}t} 
\end{displaymath}

At early times, and for low-mass haloes, the formal cooling radius can be much
larger than the virial radius. In this case, the infalling gas is never
expected to come to hydrostatic equilibrium, and the supply of cold gas for star
formation is limited by the infall rate. In this `rapid cooling regime', we
assume that all new diffuse gas that is accreted onto the halo is immediately
made available for star formation in the central galaxy of the halo under
consideration. 

In its standard implementation, the Munich model follows explicitly dark matter
haloes when they are accreted onto larger systems. This allows the dynamics of
satellite galaxies residing in infalling structures to be properly followed,
until their parent dark matter substructures are completely destroyed by tidal
truncation and stripping \citep[e.g.][]{DeLucia_etal_2004}. When this happens,
galaxies are assigned a residual surviving time that is estimated from the
relative orbit of the two merging objects, at the time of subhalo disruption,
using the following implementation of the Chandrasekhar dynamical friction
formula:
\begin{equation}
  \tau_{\rm merge}\,=\,f_{\rm fudge} \, \frac{1.17}{\ln \Lambda_{\rm df}} \, 
\frac{D^2}{\rvir^2} \, \frac{M_{\rm main}}{M_{\rm sat}}\,\tdyn
\label{eq:mergmunich}
\end{equation}
In the above equation, $D$ is the distance between the merging halo and the
centre of the structure on which it is accreted, $\rvir$ is the virial radius
of the accreting halo, $M_{\rm sat}$ is the (dark matter) mass associated with
the merging satellite, and $M_{\rm main}$ is the (dark matter) mass of the
accreting halo. The dynamical time of the halo, $\tdyn$, is given by
\begin{equation}
\tdyn = \frac{\rvir}{\vvir} = \left(\frac{\rvir^3}{G \,\mvir}\right)^{1/2}\,.
\label{eq:tdyn}
\end{equation}
Note that with the definition of virial mass adopted here, $\tdyn=0.1/H(z)$ and
is independent of the halo mass.  The Coulomb logarithm $\Lambda_{\rm df}$ is
taken to be $1+M_{\rm main}/M_{\rm sat}$.

For the purposes of this analysis, we have used Eq.~\ref{eq:mergmunich} also
for FOF-based merger trees, without adding any additional orbital dependence,
and using $\rvir$ in place of $D$. As in \citet{DeLucia_and_Blaizot_2007}, we
have assumed $f_{\rm fudge}=2$. This was originally introduced to reduce the
slight excess of bright galaxies otherwise produced, and was motivated by some
preliminary work by \citet{Springel_etal_2001}, who noted that merging times
inferred from Eq.~\ref{eq:mergmunich} are typically shorter than those directly
measured using higher resolution numerical simulations. As mentioned in Section
\ref{sec:intro}, more recent work has confirmed that the classical dynamical
friction formula tends to under-estimate merging times
\citep{Boylan-Kolchin_Ma_Quataert_2008,Jiang_etal_2008}.  While these studies
have not yet converged on the proper adjustment(s) to the classical dynamical
friction formulation, they suggest that the appropriate correction(s) cannot be
absorbed in a fudge factor in front of Eq.~\ref{eq:mergmunich}. For example,
while the fudge factor adopted in \citet{DeLucia_and_Blaizot_2007} is in good
agreement with findings from \citet{Boylan-Kolchin_Ma_Quataert_2008} for
$M_{\rm sat}/M_{\rm host}\approx 0.1$, it is significantly lower than what
found for smaller mass ratios.

\subsection{The Durham model}

The version of the Durham model used in this study is described in
\citet{Bower_etal_2006}. For full details on the cooling and merging times
modelling, the reader is referred to \citet{Benson_etal_2003} and
\citet{Cole_etal_2000}. A few small modifications were necessary in order to
make the model run on the merger trees described in Section
\ref{sec:simstrees}. These are briefly discussed below.

The hot gaseous component in dark matter haloes is assumed to have a density
profile described by the $\beta$-model:
\begin{displaymath}
\rho_{\rm g}(r) = \frac{\rho_0}{[1+(r/r_{\rm core})^2)]^{3\beta/2}}
\end{displaymath}
where $\rho_0$ is the density at the centre of the halo, $r_{\rm core}$ is the
radius of the core, and $\beta$ is a parameter that sets the slope of the
profile at radii larger than $r_{\rm core}$. The model assumes a gas density
profile with $r_{\rm core} = 0.07 \cdot \rvir$ and $\beta=2/3$ for all haloes,
in absence of energy injection \citep{Benson_etal_2003}. Since feedback is
switched off in this study, these are the model parameters adopted for our
comparison. The temperature profile of the gas is assumed to be isothermal at
the virial temperature. 

A set of `halo formation events' are defined throughout each merger tree, and
cooling calculations are begun and reset at these formation events. In
particular, each halo with no progenitor is flagged as a `formation event'. For
all other haloes in the tree, a `formation event' flag is assigned to all those
for which their mass is equal to, or larger than, twice the mass of the halo at
the previous formation event in that branch. For these haloes then, formation
events correspond to halo mass doublings. 

At each new snapshot, the mass of gas that falls onto the halo is given by:
\begin{displaymath}
  M_{\rm infall} = {\rm max}\left[(\mvir - \sum_i \mvir^{i}) \cdot 
    f_{\rm b}, 0.0\right] 
\end{displaymath}
where the sum extends over all the progenitors of the halo under
consideration. Although $M_{\rm infall}$ is accumulated for each snapshot, it
is only added to the hot gas of the halo at a formation event. 

As for the Munich model, the following calculations are then done using $20$
time-steps between each pair of simulation snapshots. A local cooling time is
defined using Eq.~\ref{eq:tcool}, and the metal free cooling function computed
from {\small CLOUDY} v8.0 \citep{Ferland_etal_1998}. A cooling radius is then
computed by equating the local cooling time to the time since the last
formation event (the cooling radius is not allowed to exceed the virial
radius). An infall radius is also computed, as the minimum between the cooling
radius and the free-fall radius (i.e. the radius within which gas has had time
to fall ballistically to the halo centre, assuming that it began at rest at the
previous formation event). The mass of gas that infalls onto the central galaxy
during each time-step is finally computed as the difference between the hot gas
enclosed within the current infall radius and the mass that was inside the
infall radius at the previous time-step.

Galaxy mergers are treated in a way similar to that done in the Munich model,
with a few differences. When a new halo forms, each satellite is assumed to
enter the halo on a random orbit (the most massive pre-existing galaxy becomes
the central galaxy of the remnant halo). The dynamical friction formulation
adopted in the Durham model is given in \citet{Cole_etal_2000} and reads as:
\begin{equation}
\tau_{\rm merge} = f_{\rm fudge}\, \Theta_{\rm orbit}\, 
\frac{0.3722}{\ln \Lambda_{\rm df}}\, \frac{M_{\rm main}}{M_{\rm
    sat}} \,\tdyn
\label{eq:mergdurham}
\end{equation}
where $M_{\rm halo}$ is the mass of the halo in which the satellite orbits, and
$M_{\rm sat}$ is the mass of the satellite galaxy including the mass of the
dark matter halo in which it formed. The Coulomb logarithm $\Lambda_{\rm df}$
is taken to be $M_{\rm halo}/M_{\rm sat}$. The orbital dependence is contained
in $\Theta_{\rm orbit}$, modelled as a log-normal distribution with mean $<\log
\Theta_{\rm orbit}> = -0.14$ and dispersion $<(\log\Theta_{\rm orbit} - <\log
\Theta_{\rm orbit}>)^2>^{1/2} =0.26$. \citet{Bower_etal_2006}, and the
stripped-down version used in this study, assume the value $1.5$ for the
dimensionless parameter $f_{\rm fudge}$. Merger times are reset at each
formation event, re-extracting orbital parameters for each satellite.

\subsection{MORGANA}
\label{sec:morgana}

All details about the modelling of gas cooling and galaxy mergers adopted in
{\small MORGANA} can be found in \citet*{Monaco_Fontanot_Taffoni_2007}, and we
refer to this paper for full details. We note that in its original formulation,
{\small MORGANA} uses merger trees obtained using {\small PINOCCHIO}
\citep{Monaco_etal_2002}. This algorithm, based on Lagrangian perturbation
theory, has been shown to provide mass assembly histories of dark matter haloes
that are in excellent agreement with results from numerical simulations
\citep{Li_etal_2007}. For the purposes of this study, we have adapted {\small
  MORGANA} to run on the numerical merger trees described in Section
\ref{sec:simstrees}. This required some small modifications that are described
below. We note that {\small PINOCCHIO} does not provide information on dark
matter substructures, so {\small MORGANA} is essentially based on FOF merger
trees. 

The hot halo phase is assumed to be spherical, in hydrostatic equilibrium in a
NFW halo, described by a polytropic equation of state with index $\gamma_p =
1.15$, and is assumed to fill the volume between the cooling radius and the
virial radius of the halo, where it is pressure balanced. The equilibrium
configuration of the hot halo gas is computed at each time-step, assuming that
the gas re-adjusts quasi-statically to the new equilibrium configuration, in
absence of major mergers. Under these assumptions, one obtains:
\begin{displaymath}
\rho_{\rm g}(r) = \rho_{g0} \left(1-a\left(1-\frac{\ln (1+c_{\rm nfw}x)}{c_{\rm
    nfw}}\right)\right)^{1/(\gamma_p -1)}
\end{displaymath}
\begin{displaymath}
T_{\rm g}(r) = T_{g0}\left(1-a\left(1-\frac{\ln (1+c_{\rm nfw}x)}{c_{\rm
    nfw}}\right)\right)
\end{displaymath}
where
\begin{displaymath}
a = \frac{3 (\gamma_p-1)}{\eta_0 \gamma_p} \left( \frac{c_{\rm nfw} (1+c_{\rm
nfw})}{(1+c_{\rm nfw}) \ln(1+c_{\rm nfw})-c_{\rm nfw}} \right)
\end{displaymath}
In the above equations, $c_{\rm nfw}= r_{\rm halo}/r_s$, and $x=r/r_s$, where
$r_s$ is the scale radius of the NFW halo. The constants $\rho_{g0}$ and
$T_{g0}$ are defined as the extrapolations to $r=0$ of the density and
temperature profile, while $\eta_0$ is the extrapolation to $r=0$ of the
function $\eta(r) = T_{\rm g}(r)/T_{\rm vir}$. The halo virial temperature
$T_{\rm vir}$ is defined as $1/3 \mu m_{\rm H} \vvir^2/k$.

At each new snapshot, the mass of gas that falls onto the halo is computed
as:
\begin{displaymath}
 M_{\rm infall} = f_{\rm b} \cdot {\rm max} \left[ \mvir - (\sum_i
   M_{\rm}^i + M^{\rm max}_{\rm vir}) , 0.0 \right]
\end{displaymath}
where the sum extends over all progenitors of the halo that are not its main
progenitor, and $M^{\rm max}_{\rm vir}$ is the maximum virial mass of the main
progenitor, considering all previous snapshots. The infall rate of new gas is
assumed to be constant over the time interval between each pair of simulation
snapshots. The equations below are then solved using a Runge-Kutta integrator
with adaptive time-steps. 

The cooling rate of a shell of gas of width $\Delta r$, at a radius $r$, is
computed as:
\begin{displaymath}
\Delta \left( \frac {dM_{\rm cool} (r)}{dt} \right) = \frac{4 \pi r^2 \rho_{\rm
    g}(r)\Delta r}{t_{\rm cool}(r)}
\end{displaymath}
where $t_{\rm cool}$ is the local cooling time, computed as in
Eq.~\ref{eq:tcool}.  As in the Munich model, cooling rates are computed using
the collisional ionization cooling curves by
\citet{Sutherland_and_Dopita_1993}, with primordial composition. The mass
deposition rate is then computed by summing up the contributions from all mass
shells. The summation is carried our by taking into account the radial
dependence of the gas density, and provides the following total mass deposition
rate:
\begin{displaymath}
\frac{dM_{\rm cool}}{dt} = 
\end{displaymath} 
\begin{displaymath}
\hspace{0.5cm}  \frac{4 \pi r_s^3 \rho_{g0}}{t_{{\rm cool,}0}}
\int_{r_{\rm cool}/r_s}^{c_{\rm nfw}}\left[1-a\left(1 -
  \frac{\ln(1+t)}{t}\right)\right]^{2/(\gamma_p-1)} t^2 dt
\end{displaymath}
where $t_{\rm cool,}0$ is computed using the central density $\rho_{g0}$ and
the temperature of the gas at $r_{\rm cool}$.

Finally, by equating the mass cooled in a time interval $dt$ with the mass
contained in a shell $dr$, one obtains the evolution of the cooling radius,
that is treated as a dynamical variable in this model:
\begin{displaymath}
\frac{dr_{\rm cool}}{dt} = \frac{dM_{\rm cool}/dt}{4 \pi \rho_{\rm g}(r_{\rm
    cool}) r_{\rm cool}^2} - c_{\rm s}
\end{displaymath}
where $c_{\rm s}$ is the sound speed computed at $r_{\rm cool}$, and is added
to the right hand side of the above equation to allow the `cooling hole' to
close at the sound speed.

The cooling calculation is started when a halo appears for the first time, with
$r_{\rm cool} = 0.001 \cdot r_s$, and is reset after any halo major merger
($M_{\rm sat}/M_{\rm main} > 0.2$). Finally, cooled gas is incorporated onto
the central galaxy at the following rate:
\begin{displaymath}
\frac{dM_{\rm cold}}{dt} = \frac{M_{\rm cooled}}{n_{\rm dyn} \tdyn(r_{\rm
    cool})}
\end{displaymath}
where $n_{\rm dyn}$ is treated as a free parameter (we assume a value of 0.3
for this parameter, as in the standard {\small MORGANA} model) and represents
the number of dynamical times, computed at the cooling radius, required for the
cooled gas to be incorporated onto the central galaxy. In the comparison
discussed below, the above delay in the incorporation of the cooled gas is
neglected.

When a halo is accreted on a larger structure, orbital parameters are extracted
randomly from suitable distributions that are based on results from numerical
simulations. In particular, the eccentricity of the orbit ($\epsilon = J/J_c$,
where $J$ is the initial angular momentum of the orbit and $J_c$ is the angular
momentum of a circular orbit with the same energy) is extracted from a Gaussian
with mean 0.7 and variance 0.2, while the energy of the orbit ($x_c=r_c/r_h$,
where $r_h$ is the halo radius and $r_c$ is the radius of a circular orbit with
the same energy) is assumed to be 0.5 for all orbits. To model galaxy mergers,
\citet{Monaco_Fontanot_Taffoni_2007} use a slight update of the formulae
provided by \citet{Taffoni_etal_2003}, that take into account dynamical
friction, mass loss by tidal stripping, tidal disruption of subhaloes, and
tidal shocks. Galaxy merger times are computed interpolating between the case
for a `live satellite' (the object is subject to significant mass losses) and
that of a `rigid' satellite (no mass loss):
\begin{displaymath}
\tau_{\rm merge,live} = \frac{\tdyn}{f_{\rm sat}}\,
\left(\xi_1 f_{\rm sat}^{0.12} + \xi_2 f_{\rm sat}^{2} \right)
\end{displaymath}
\begin{displaymath}
\hspace{0.5cm} \times \left( 0.25 \,f_{\rm nfw}^{-6} +0.07 f_{\rm nfw} +1.123
\right) \left( 0.4 + \xi_3 (\epsilon-0.2) \right)
\end{displaymath}
\begin{equation}
\tau_{\rm merge,rigid} = 0.46 \, \frac{\tdyn}{f_{\rm sat}}\,
\cdot \left( 1.7265 + 0.0416 \, c_{\rm nfw} \right)
\frac{x_c^{1.5}}{\ln\Lambda_{\rm df}}
\label{eq:mergmorgana}
\end{equation}
In the above equations, $f_{\rm sat} = M_{\rm sat}/\mvir$ and $f_{\rm
  nfw} = c_{\rm sat}/c_{\rm nfw}$. $M_{\rm sat}$ and $c_{\rm sat}$ are the
corresponding virial quantities for the satellite halo, and $\xi_i$ are
polynomial functions of $x_c$, whose expressions can be found in appendix A of
\citet{Monaco_Fontanot_Taffoni_2007}.  The Coulomb logarithm is given by
$\Lambda_{\rm df}=1+1/f_{\rm sat}$.  Merger times are reset after each halo
major merger, re-extracting orbital parameters for each satellite galaxy.

\subsection{Model differences and similarities}
\label{sec:moddiff}

The previous sections illustrate that the implementations of gas cooling and
galaxy mergers in the three models used in this study differ in a number of
details. Both the Durham and the Munich models adopt variations of the original
cooling model proposed by \citet{White_and_Frenk_1991}, but they use different
gas profiles, and different definitions of the `cooling time' (used to
calculate the cooling radius). In addition, the cooling calculation is reset in
the Durham model at each `formation event', and the Munich model assumes very
efficient cooling in the rapid cooling regime. Finally, cosmological infall of
gas onto the halo occurs at a constant rate between each pair of snapshots in
the Munich model, while occurs at formation events for the Durham model. A
cored density profile, like the one adopted in the Durham model, is expected to
give lower cooling rates than an isothermal profile. It is not clear, however,
how this naive expectation is affected by the other different assumptions
discussed above.

The adopted modelling for merger times is also very similar in the Durham and
Munich models, as well as for the `rigid' case in the {\small MORGANA} model.
Indeed, by comparing Eqs.~\ref{eq:mergmunich}, \ref{eq:mergdurham}, and
\ref{eq:mergmorgana} with $D=\rvir$ in the Munich model, $\Theta_{\rm orbit}=1$
in the Durham model, and $c_{\rm NFW}=10$ in (the rigid version of) {\small
  MORGANA}, the implementations differ only in the numerical pre-factor
(Munich: $1.17 \,f_{\rm fudge}$; Durham: $0.3722 \,f_{\rm fudge}$; {\small
  MORGANA}: 0.348) and in the implementation of the Coulomb logarithm.  (The
full `live' version of dynamical friction in {\small MORGANA} is somewhat more
complicated.) 

{\small MORGANA} differs significantly from the Durham and the Munich models,
both in its gas cooling and merger times implementations. As explained above,
in the Durham and Munich model the cooling radius is computed by equating the
local cooling time (Eq.~\ref{eq:tcool}) to some given time (the halo dynamical
time in the Munich model and the time since the last formation event in the
Durham model). {\small MORGANA} computes the cooling rate at each mass shell,
integrates over the contribution of all shells, and follows the evolution of
the cooling radius by assuming that the transition from the hot to the cold
phase is fast enough to create a sharp edge in the density profile of the hot
gas. The cooling radius `closes' at the sound speed. 

As mentioned in Section \ref{sec:intro}, \citet{Viola_etal_2008} have shown
that results from this model are in good agreement with controlled
hydrodynamical simulations of isolated haloes, with hot gas in hydrostatic
equilibrium in a NFW halo. Viola et al. have also shown that their
implementation of the classical model underpredicts the amount of gas cooling
with respect to the model adopted in {\small MORGANA}, particularly at early
times. Their implementation of the `classical' model differs in detail from
those adopted in the Munich and Durham models, however. In addition, it is
unclear that results obtained from their controlled simulations should remain
valid when using cosmologically motivated halo merger histories, as done in
this study. The formulation adopted to model dynamical friction in {\small
  MORGANA} is based on controlled simulations and analytic models, and takes
into account dynamical friction, mass loss by tidal stripping, tidal disruption
of subhaloes, and tidal shocks. Since the other two models treat satellites as
rigid systems, they are expected to provide somewhat shorter merger times. We
will come back to a more detailed comparison between the adopted formulations
in Section \ref{sec:mergtimes}.

{\small MORGANA} and the Munich model compute cooling rates using results by
\citet{Sutherland_and_Dopita_1993}, while the Durham model adopts updated rates
computed using {\small CLOUDY}. We have verified that the metal free cooling
function computed using {\small CLOUDY} does not differ significantly from the
primordial cooling function from \citet{Sutherland_and_Dopita_1993}. It should
be noted, however, that this is not the case for non-zero metallicities, where
the Sutherland \& Dopita calculation tends to over-estimate the cooling rates
computed using {\small CLOUDY}.

Finally, we note that in all three models used in this study, gas cooling
occurs only on central galaxies. When galaxies are accreted onto a more massive
system, their hot reservoir is assumed to be instantaneously stripped and
associated with the parent halo of the remnant central galaxy.

\section{Milky-Way haloes}
\label{sec:mw}

\begin{figure*}
\bc
\resizebox{17cm}{!}{\includegraphics[]{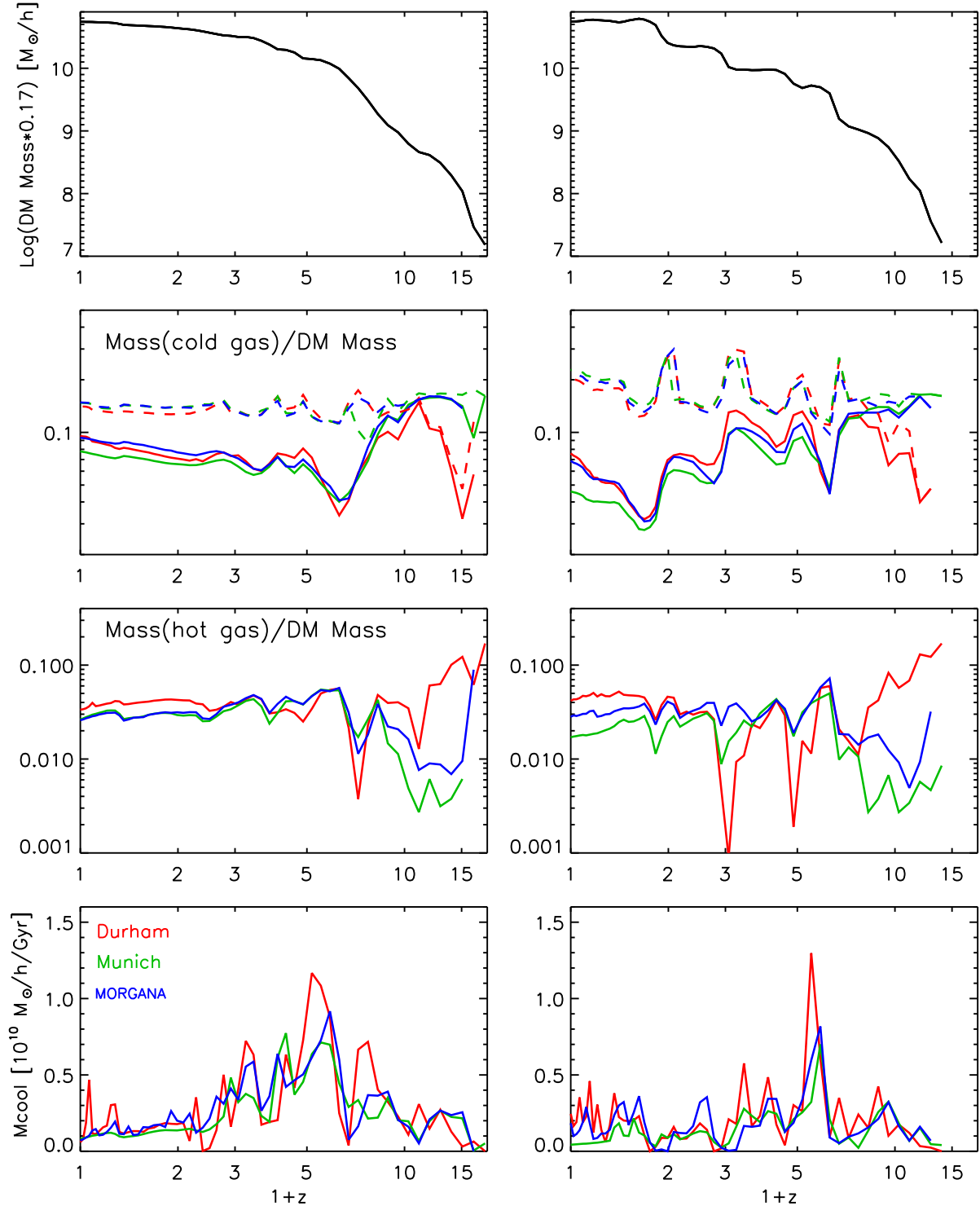}}
\caption{From top to bottom: dark matter, cold gas, hot gas, and gas cooling
  rate for two MW-like haloes. Only quantities associated with the central
  galaxy and its main progenitors are plotted. Red lines show results from the
  Durham model, green lines show results from the Munich model and blue lines
  are for {\small MORGANA}. Solid and dashed lines in the second panels from
  the top show results for the $t_{mrg}=\infty$ and $t_{mrg}=0$ run,
  respectively. The haloes shown in this figure provide two representative
  examples for a halo with a rather quiet mass accretion history (left panels)
  and a halo with a larger number of massive mergers (right panels).}
\label{fig:mwexamples}
\ec
\end{figure*}

In this section we will analyse results of the three models described above for
a sample of Milky-Way like haloes. As discussed in Section \ref{sec:simstrees},
these haloes have been selected only on the basis of their present virial mass,
with no additional constraint on their merger history or isolation.  An
in-depth analysis of the full sample of Milky Way-mass haloes in the MS-II
(comprised of over 7600 haloes) is presented in
\citet{Boylan-Kolchin_etal_2010}.

Fig.~\ref{fig:mwexamples} shows the dark matter mass scaled by the universal
baryon fraction (top panel), the evolution of the cold gas (second panels from
top) and hot gas (third panels from top) components associated with the central
galaxy, and the cooling rate (bottom panels) for two MW-like haloes. The cold
and hot components have been normalized by the dark matter mass of the parent
halo, which is the same in all three models by construction. Red, green and
blue lines show results from the Durham, Munich and {\small MORGANA} models,
respectively. The haloes chosen for this figure provide two representative
examples for a MW halo with a rather quiet mass accretion history (left panels)
and one with a larger number of massive mergers (right panels). To avoid
complications arising from a different treatment of merging times, the
evolution of the cold gas content is shown for both $t_{mrg}=\infty$ (solid
lines) and $t_{mrg}=0$ (dashed lines). The amount of hot gas associated with
the central galaxy, as well as the cooling rate, are not affected by the
particular merging model adopted, because the hot gas associated with galaxies
falling onto a larger system (i.e. becoming satellites) is instantaneously
stripped and associated with the hot gas component of the main halo, in all
three models.

Fig.~\ref{fig:mwexamples} shows a quite good degree of agreement between the
three models used in this study. At high redshift, when the halo is in a rapid
accretion regime, a large fraction of its baryonic mass is converted into cold
gas in the Munich model and in {\small MORGANA}, while cooling appears to be
less efficient in this regime in the Durham model. By redshift $\sim 7-9$, all
models converge to about the same cold gas and hot gas fractions. At lower
redshift, the evolution of both components is very similar, with small
differences between different models at present (see below). As expected, the
evolution of both baryonic components is more noisy for the halo whose mass
accretion history is characterized by a larger number of important mergers. In
particular, the Durham model shows a quite noisy behaviour in the hot gas
evolution, which is due to the fact that gas infalling onto the halo is added
to the hot gas component only at `formation events' (see Section
\ref{sec:models}). For the halo shown in the right panels, the differences
between the amounts of cold and hot gas predicted by the three models are
somewhat larger than for the halo with a smoother accretion history shown in
the left panels, and differences appear to accumulate at each merger event.

The predicted cooling rates are very noisy for all three models used in this
study, and in both examples shown in Fig.~\ref{fig:mwexamples}. In these, the
highest cooling rates are obtained in the Durham model and are of the order of
$\sim 1.8\times 10^{10} \,{\rm M}_{\odot} {\rm Gyr}^{-1}$. In a number of the
other haloes in the MW sample, cooling rates as high as about twice this value
are obtained, and there is some tendency for the Durham model to provide the
noisiest behaviour, in particular at very low redshift ($z<0.5$). Overall,
however, the evolution of the cooling rates predicted by the three models is
quite similar.

\begin{figure*}
\bc
\resizebox{15.5cm}{!}{\includegraphics[angle=90]{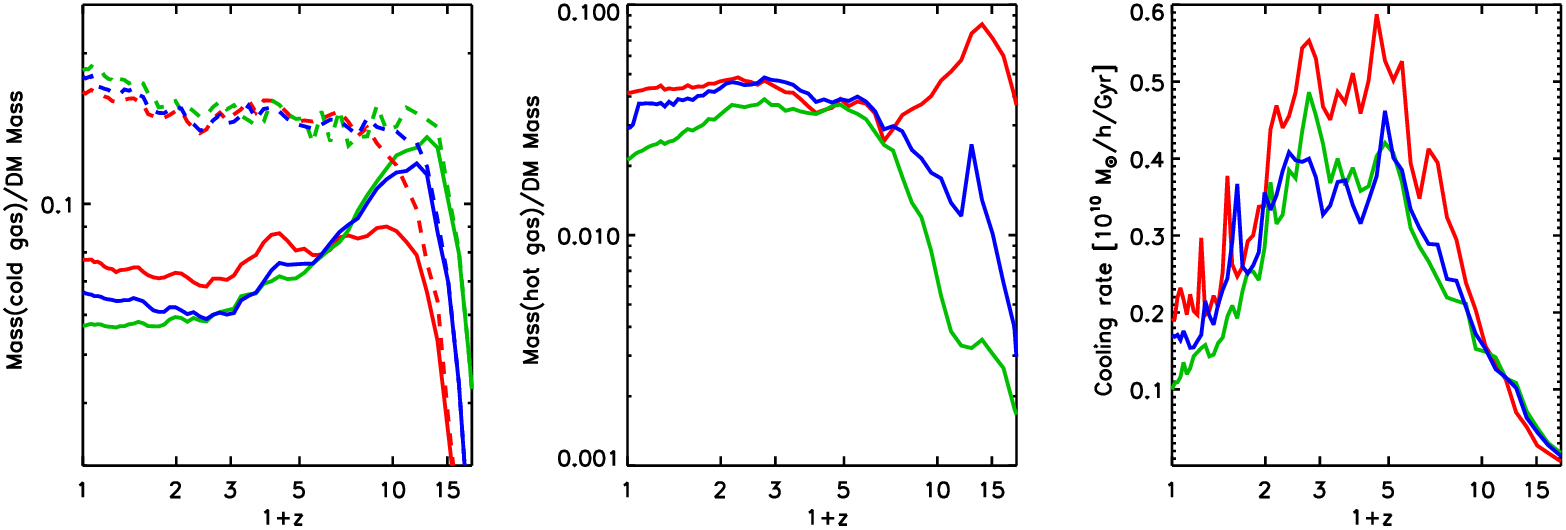}}
\caption{From left to right: cold gas, hot gas, and gas cooling between two
  subsequent snapshots. Each line shows the average computed over 100 MW-like
  haloes. Solid and dashed lines in the left panels show results for the
  $t_{mrg}=\infty$ and $t_{mrg}=0$ run, respectively. Colour-coding is as in
  Fig.~\ref{fig:mwexamples}.}
\label{fig:mwavg}
\ec
\end{figure*}

\begin{figure*}
\bc
\resizebox{16cm}{!}{\includegraphics[angle=90]{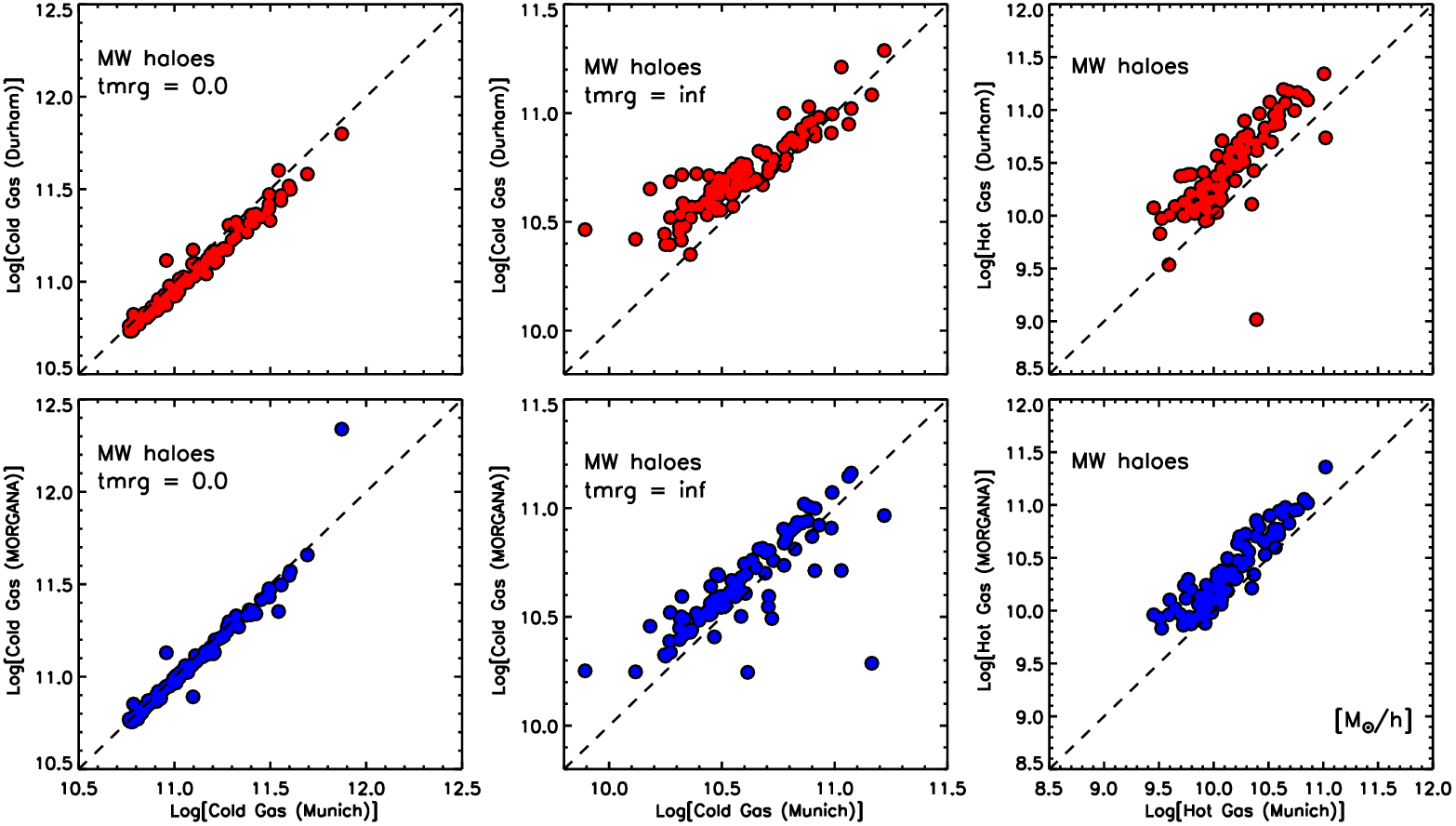}}
\caption{Cold gas for the $t_{mrg}=0$ case (left panels), cold gas for the
  $t_{mrg}=\infty$ case (middle panels) and hot gas (right panels) associated
  with the central galaxy for the MW sample. Top panels show results obtained
  by the Durham code against those from the Munich model. Bottom panels show
  results by the {\small MORGANA} code against those from the Munich model. The
  dashed line is the one-to-one relation and is plotted to guide the eye.}
\label{fig:mwcoldhot}
\ec
\end{figure*}

In order to study any systematics in the evolution predicted by the three
models used in this study, we have run each code on the total sample of 100
MW-like haloes. Fig.~\ref{fig:mwavg} shows the average evolution of the cold
(left panel) and hot (middle panel) gas components associated with the central
galaxy, and of the cooling rate (right panel). In the left panel, solid and
dashed lines correspond to the $t_{mrg}=\infty$ and $t_{mrg}=0$ run,
respectively. Red, green and blue lines show again results from the Durham,
Munich and {\small MORGANA} models, respectively. The fraction of cold gas
associated with the central galaxy rises steeply in all models and at
progressively decreasing redshift for the Munich model, {\small MORGANA}, and
for the Durham model. The later evolution of this component is very similar in
all three models when merger times are set to zero, with a slight tendency for
the Munich model to predict the largest cold gas fraction and the lowest hot
gas fraction at present (see below). When merger times are set to $\infty$, the
Munich model predicts the highest cold gas fractions at early times and the
lowest cold gas fractions at low redshift. {\small MORGANA} has a similar
behaviour but the cold gas fraction rises slightly later than in the Munich
model at early times, and is slightly larger than predicted by the Munich model
at late times. Finally, in the Durham model, the cold gas fraction starts
rising at $z\sim 15$, reaches values of $\sim 0.09$ at $z\sim 10$, and stays
almost constant down to $z=0$. The evolution of the hot gas fraction reflects
that of the cold component at high redshift, with the Durham model predicting
the largest hot gas fractions, due to a less efficient cooling with respect to
the other two models. At $z<4$, the Durham model is on average still slightly
above the predictions from the Munich model and from {\small MORGANA}. On
average, the Durham model predicts the highest cooling rates in the redshift
interval $2-4$. Predictions from the Munich model and from {\small MORGANA} are
very similar, with a broad peak over the same redshift interval, and with a
rapid decline at $z<2$.

Fig.~\ref{fig:mwcoldhot} shows the amount of cold (left panels for $t_{mrg}=0$
and middle panels for $t_{mrg}=\infty$) and hot gas (right panels) for all
MW-like haloes in our sample, at $z=0$. Top and bottom panels compare the
Durham model and {\small MORGANA} with the Munich model, respectively. The
agreement between the Munich model and {\small MORGANA} is very good for the
predicted cold gas, particularly when zero merging times are adopted, with only
a very slight tendency for {\small MORGANA} to predict a smaller gas content
with respect to the Munich model. When suppressing galaxy mergers
($t_{mrg}=\infty$), the agreement is still good, but the scatter is much larger
and there is a systematic trend for larger cold gas amounts in {\small
  MORGANA}. The examples discussed above suggest that this might be due to a
different treatment of the rapid accretion regime in these two models. The same
systematic trend is visible for the hot gas content (right panel), which is
always larger in {\small MORGANA} than in the Munich model. The trends are
similar with the systematics being slightly stronger in the top panels, that
compare prediction from the Durham model to those from the Munich model. This
figure demonstrates that all three models used in this study provide very
similar predictions at present, but that there are some residual systematic
trends at late times, and the evolution at high redshift is quite different
(see Fig.~\ref{fig:mwexamples} and Fig.~\ref{fig:mwavg}).


\section{SCUBA haloes}
\label{sec:scuba}

In order to investigate how the level of agreement between different models
discussed in the previous section depends on halo mass, we have complemented
our MW-like sample with a sample of {\small SCUBA}-like haloes. As discussed in
Section \ref{sec:simstrees}, these have been selected as haloes with a number
density of $10^{-5}$ at $z\sim 2$ and with a relatively massive descendant at
$z=0$ ($M_{200}=7.8 \times 10^{13}-1.3 \times 10^{15}\,M_{\odot}$). Our
decision to use a {\small SCUBA}-like sample has been partially motivated by
previous claims that {\small MORGANA} provides a good agreement with the
observed redshift distribution and number counts of 850-$\mu$m sources because
the adopted cooling model predicts significantly larger cooling rates with
respect to the `classical' cooling model
\citep{Fontanot_etal_2007,Viola_etal_2008}. At this mass scale, we therefore
expect significant differences between predictions from {\small MORGANA} and
those from the Munich and Durham models, that both adopt different
implementations of the classical model.

\begin{figure*}
\bc
\resizebox{17cm}{!}{\includegraphics[]{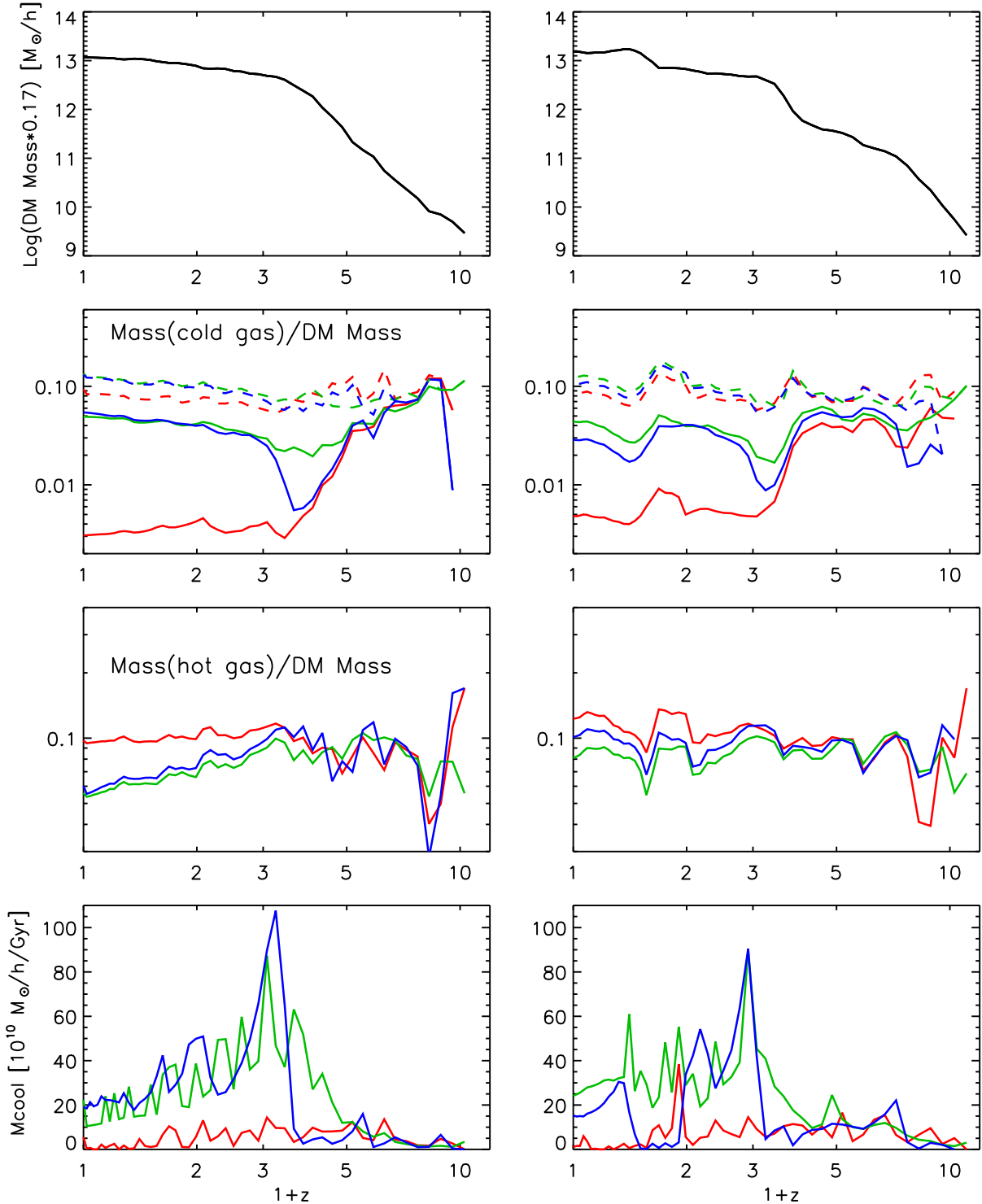}}
\caption{Same as in Fig.~\ref{fig:mwexamples}, but for two SCUBA-like haloes.}
\label{fig:scubaexamples}
\ec
\end{figure*}

Fig.~\ref{fig:scubaexamples} shows the evolution of the cold and hot components
(second and third panels from the top) normalized to the parent halo dark
matter mass, and of the cooling rates for two examples from our {\small SCUBA}
sample. As for Fig.~\ref{fig:mwexamples}, these two examples have been chosen
as representative for a halo with a rather quiet mass accretion history (left
panels), and one with a larger number of important mergers (right panels). The
dark matter mass accretion histories of the haloes under consideration are
shown in the top panels. Contrary to what expected, predictions from {\small
  MORGANA} appear to be very close to those from the Munich model, while the
Durham model deviates most from the other two, predicting systematically lower
cold gas fractions at late times. The systematics are stronger when satellite
galaxies are allowed to survive as independent entities down to $z=0$
($t_{mrg}=\infty$).

The bottom panels of Fig.~\ref{fig:scubaexamples} show that the Durham model
predicts much lower cooling rates than the Munich model and {\small MORGANA},
below $z \sim 4$. Over this redshift range, both the Munich model and {\small
  MORGANA} provide cooling rates as high as $\sim 80-100\times10^{10}\,{\rm
  M}_{\odot} {\rm Gyr}^{-1}$ in the two examples shown; values about twice (or
more) as high are obtained in a number of the other haloes from the SCUBA
sample. The behaviour predicted by the Munich model for these haloes appears to
be somewhat noisier than that predicted by {\small MORGANA}, with cooling rates
that are not, however, significantly lower. At the highest redshifts probed by
the merger trees at disposal, the hot gas fraction predicted by the Munich
model is lower than the corresponding value predicted by the Durham model and
{\small MORGANA}. As noted in the previous section, this results from a very
efficient cooling in the rapid accretion regime, as treated by the Munich
model. The evolution of the hot gas fraction is then very similar in all three
models, down to $z\sim 2$. At redshift lower than this, the Munich model and
{\small MORGANA} still provide reasonably close results, while the Durham model
is systematically higher.

\begin{figure*}
\bc
\resizebox{15.5cm}{!}{\includegraphics[angle=90]{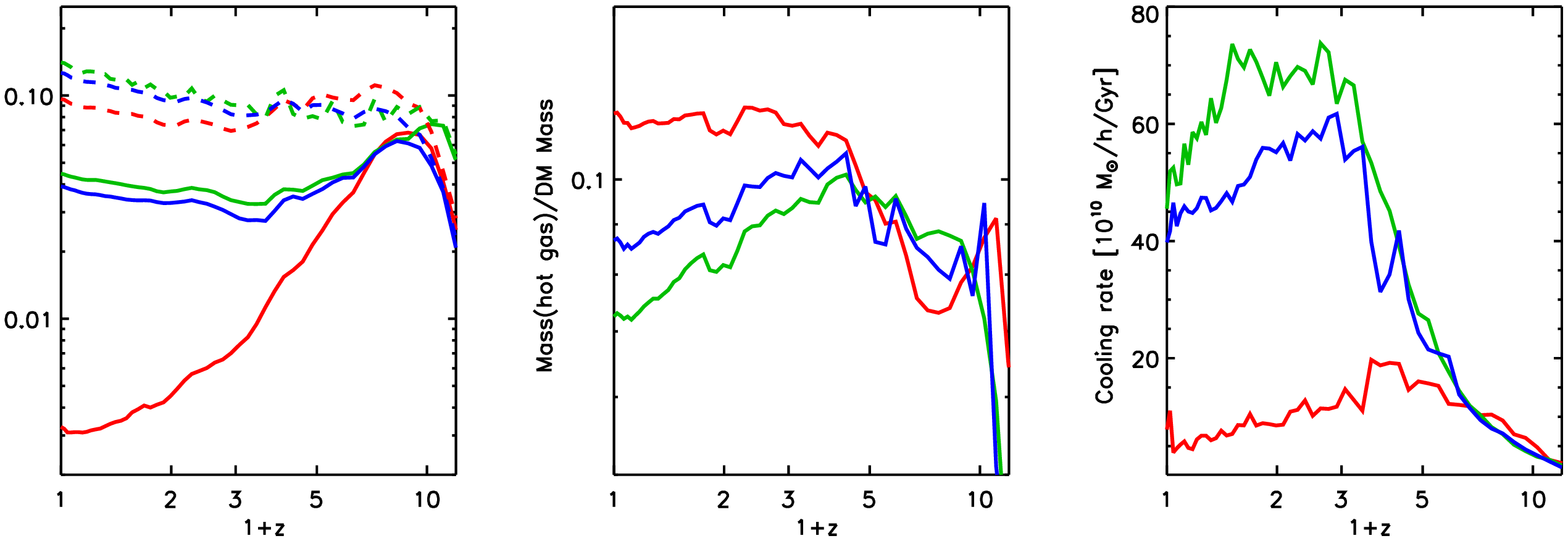}}
\caption{Same as in Fig.~\ref{fig:mwavg}, but for the SCUBA sample.}
\label{fig:scubaavg}
\ec
\end{figure*}

\begin{figure*}
\bc
\resizebox{16cm}{!}{\includegraphics[angle=90]{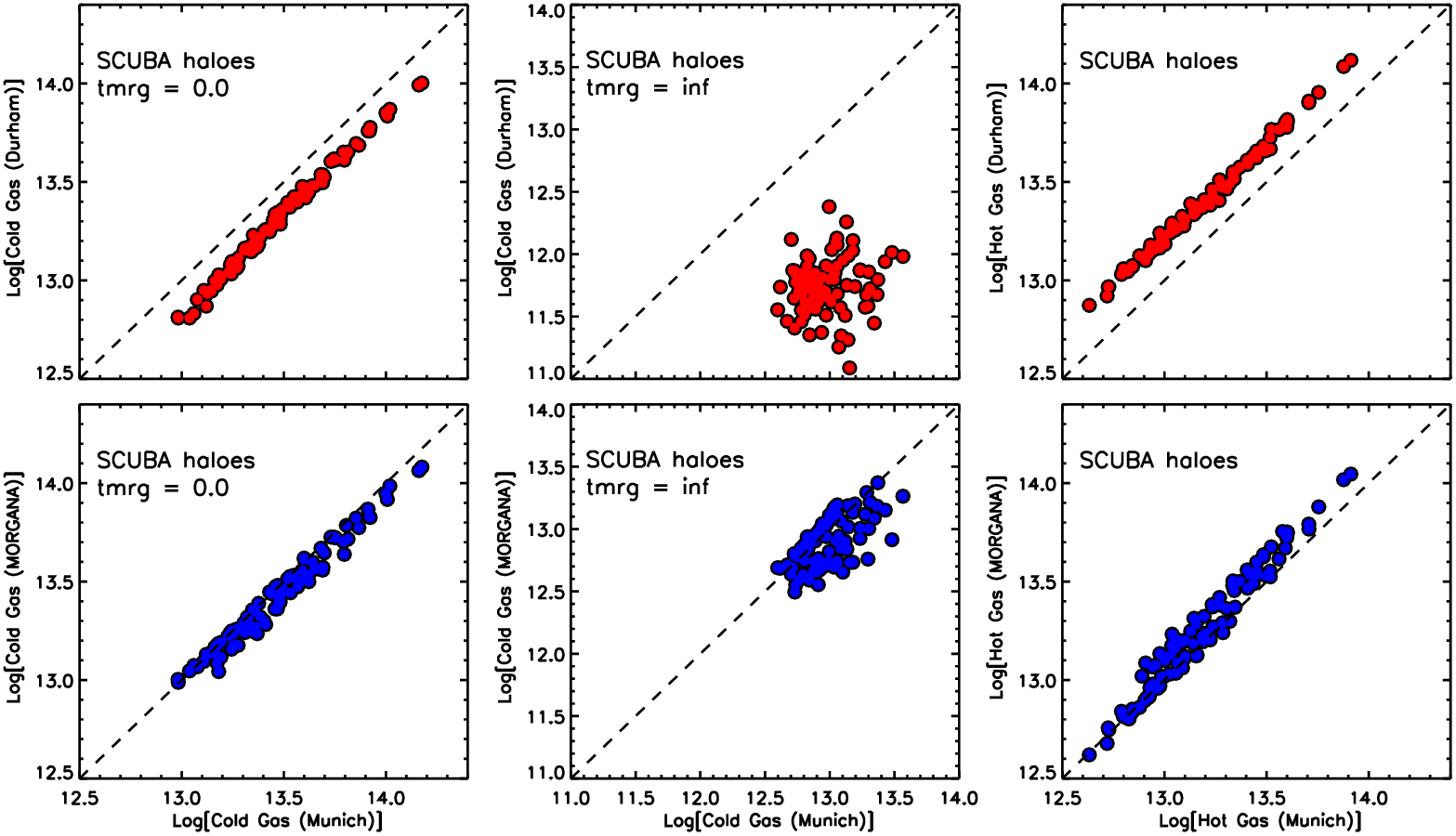}}
\caption{Same as in Fig.~\ref{fig:mwcoldhot}, but for the SCUBA sample. The
  dashed line is the one-to-one relation and is plotted to guide the eye.}
\label{fig:scubacoldhot}
\ec
\end{figure*}

The differences discussed above are clearly visible in Fig.~\ref{fig:scubaavg},
that shows the average evolution of the cold (left panel) and hot (middle
panel) gas fractions associated with the central galaxy, and of the cooling
rates, computed using all 100 haloes in the {\small SCUBA} sample. As in
previous figures, solid and dashed lines in the left panel correspond to the
$t_{mrg}=\infty$ and $t_{mrg}=0$ run, respectively. When satellite galaxies are
merged immediately after their accretion, the cold gas fractions predicted by
the Munich model and {\small MORGANA} are very similar at all redshifts, while
the Durham model tends to predict larger cold gas fractions at high redshift
and lower fractions at $z<2$. When satellite mergers are suppressed
(i.e. satellite galaxies survive as independent entities, keeping the cold gas
associated with them before accretion), the Durham model predicts significantly
lower cold gas fractions than the Munich model and {\small MORGANA}. This is
due to the significantly lower cooling rates predicted by the Durham model, as
can be seen in the right panel of Fig.~\ref{fig:scubaavg}. This panel shows
that, on average, the Munich model predicts the highest cooling rates on this
mass scale, while predictions from {\small MORGANA} are intermediate between
the Munich and Durham models. The middle panel of Fig.~\ref{fig:scubaavg} shows
the evolution of the hot gas fraction, normalized by the corresponding dark
matter mass. It shows that, as noted for the MW-like haloes, the cooling
efficiency at high redshift is highest in the Munich model. This produces the
rapid increase of cold gas visible in the left panel, and is due to a different
treatment of the rapid cooling regime in this model. At $z<2$, the Durham model
predicts the largest hot gas fraction and the Munich model the lowest, with
{\small MORGANA} again providing intermediate results.

The amount of cold and hot gas for all haloes in our {\small SCUBA} sample at
present are shown in Fig.~\ref{fig:scubacoldhot}, that compares predictions
from Durham model and from {\small MORGANA} with results from the Munich
model. From this figure, it is clear that there is a systematic trend for both
the Durham model (top panel) and {\small MORGANA} (bottom panels) to predict
lower cold gas fractions and higher hot gas fractions with respect to the
Munich model. The disagreement is stronger for the Durham model than for
{\small MORGANA}, and when galaxy mergers are suppressed. As shown above, this
is due to systematic differences in cooling rates, which appear to be more
significant for this mass scale than for MW-like haloes (compare right panels
in Fig.~\ref{fig:scubaavg} and Fig.~\ref{fig:mwavg}).

\section{Numerical resolution and subhalo schemes}
\label{sec:ressub}

In order to study how the results discussed in previous sections are affected
by numerical resolution, we have taken advantage of the mini-MSII. As explained
in Section \ref{sec:simstrees}, this simulation has been run with the same
initial conditions of the MS-II, but lower force and mass resolution (the same
as for the MS). We have identified the {\it same} haloes used in our MW-sample
and run each model on the corresponding merger trees. Haloes were matched
across the two simulations by finding, at $z=0$, all FOF groups in the
mini-MS-II within a distance of $1{\rm Mpc}\,{h}^{-1}$ of the coordinates of
the target haloes from the MS-II\footnote{This search radius is a factor of
  $\sim 7$ smaller than the typical separation between
  $10^{12}\,h^{-1}\,M_{\odot}$.}.  For each halo in the original MW sample, the
matched halo is found from this subset of haloes, as the one that minimizes the
absolute value of $\mvir($mini-MSII$)/\mvir($MS-II$)-1$.
We find that the matched halo lies within $100 {\rm kpc}\,{h}^{-1}$ of the
target halo in more than $90$ per cent of the cases. In 76 per cent of the
cases, the present virial masses differ by less than 10 per cent, while in 95
per cent of the cases matched haloes have virial masses that differ by less
than 20 per cent at present.

\begin{figure}
\bc
\resizebox{7.5cm}{!}{\includegraphics[]{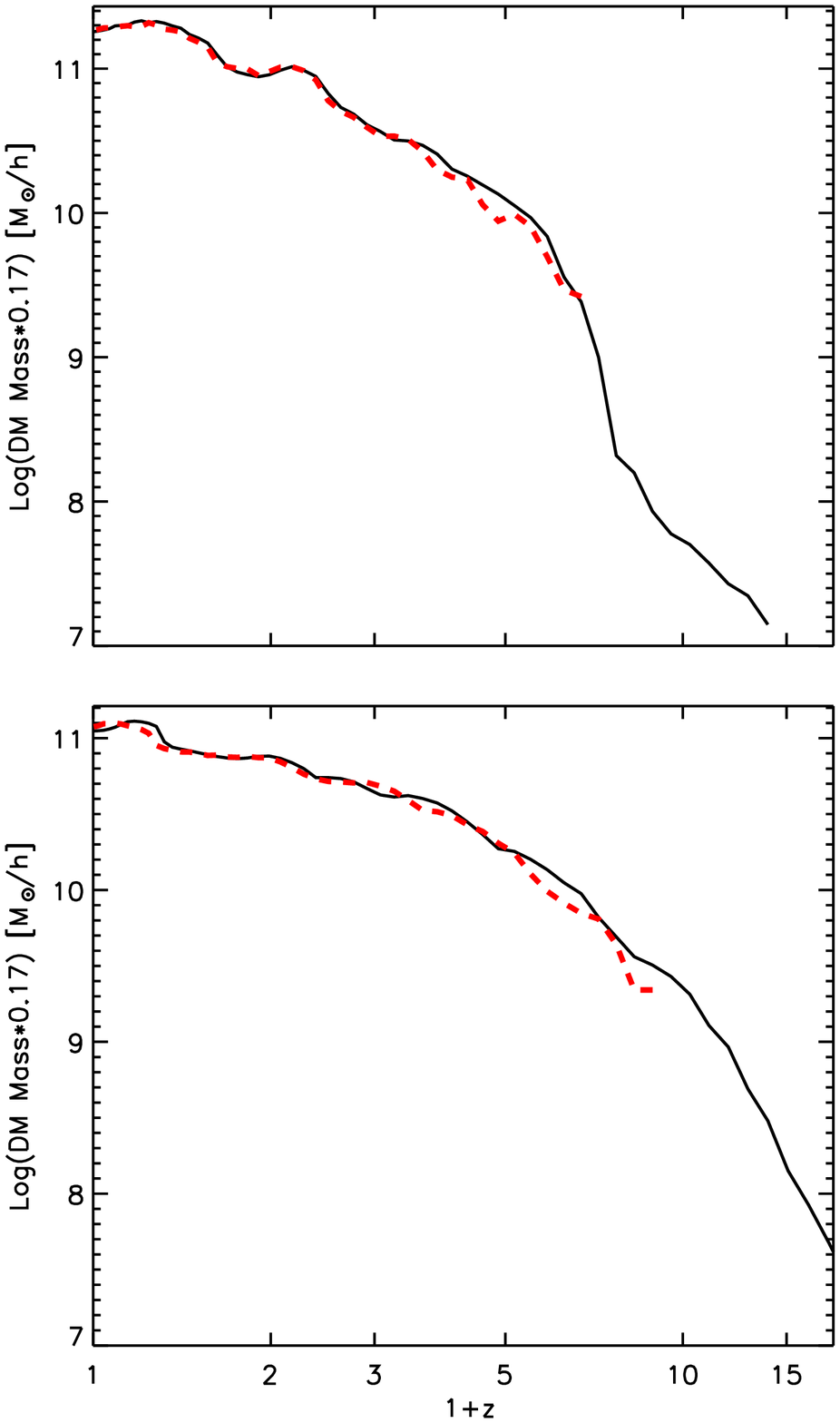}}
\caption{Mass accretion history of two of the haloes from the MW-like sample
  used in this study, at two different resolution levels. Solid lines
  show the histories constructed using the MS-II simulation, while dashed
  lines show the corresponding histories from the mini-MSII.}
\label{fig:exminiDM}
\ec
\end{figure}

Fig.~\ref{fig:exminiDM} shows the mass accretion histories of two MW-like
haloes as obtained from the MS-II (solid lines) and from the mini-MSII
(dashed lines). As for previous figures, mass accretion histories have
been constructed by connecting each halo to its main progenitor (i.e. the most
massive) at each node of the tree. At mini-MSII resolution, it is not possible
to follow the evolution of these haloes past $z\sim 7-9$, while the first
progenitors of the MW-like haloes under consideration are identified at $z\sim
14-15$ at the resolution of the MS-II.  Over the redshift range where haloes
are identified in both simulations, the mass accretion histories extracted from
the MS and mini-MSII are in quite good agreement. Most of the differences
discussed below, should then be ascribed to the ability to follow the evolution
of the parent dark matter haloes up to higher redshift in the higher resolution
simulation.

\begin{figure*}
\bc
\resizebox{15.5cm}{!}{\includegraphics[]{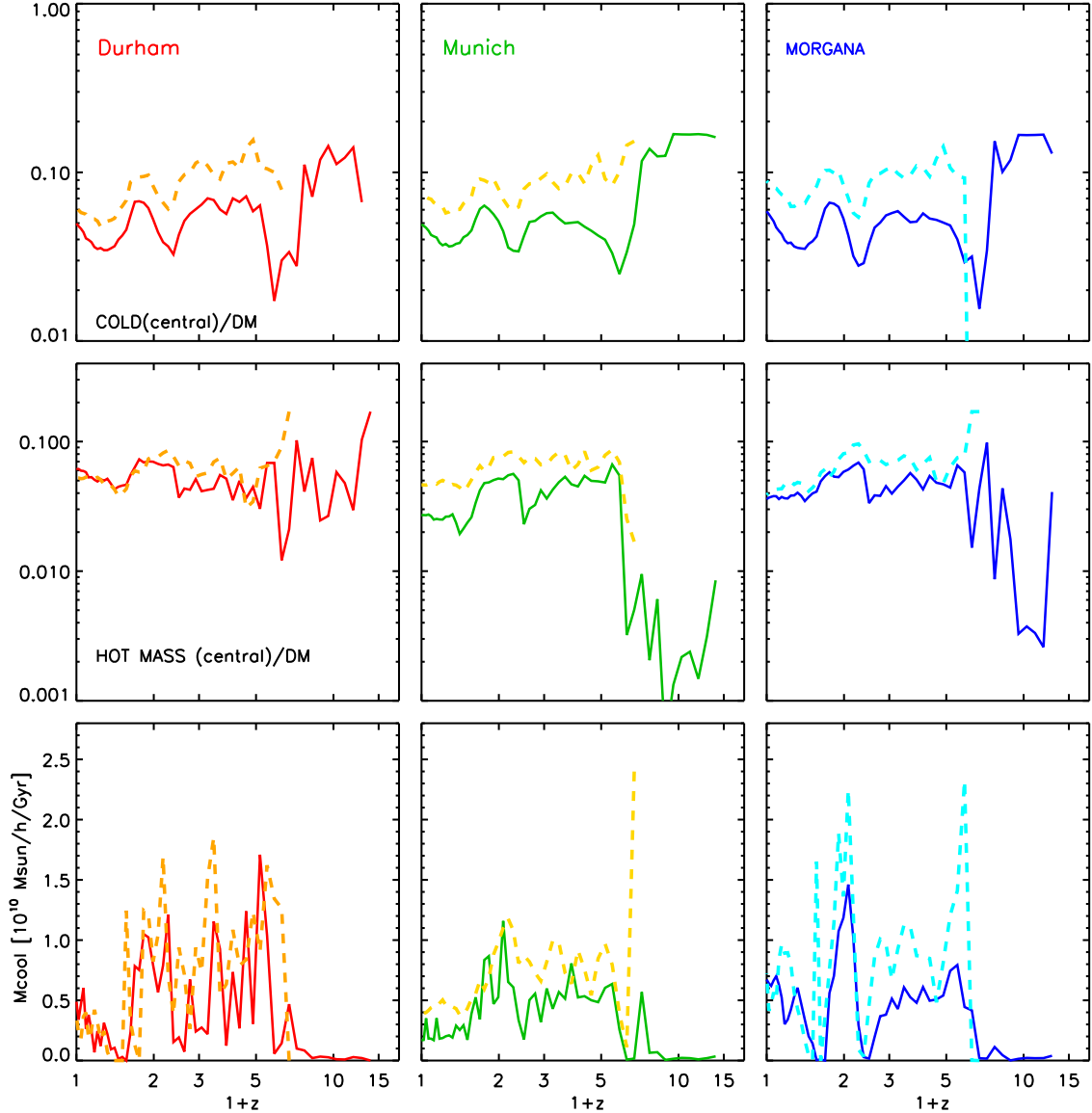}}
\caption{Evolution of the cold gas content (top panels), hot gas (middle
  panels) and the cooling rates (bottom lines) for one of the MW-like haloes
  used in this sample, at two different resolution levels. In each panel,
  solid lines show the evolution computed using the merger trees extracted
  from the MS-II, while dashed lines correspond to merger trees from
  mini-MSII. Left, middle and right panels show results from the Durham,
  Munich and {\small MORGANA} model, respectively. The mass accretion history
  of this halo is shown in the top panel of Fig.~\ref{fig:exminiDM}.}
\label{fig:exmini1}
\ec
\end{figure*}

\begin{figure*}
\bc
\resizebox{15.5cm}{!}{\includegraphics[]{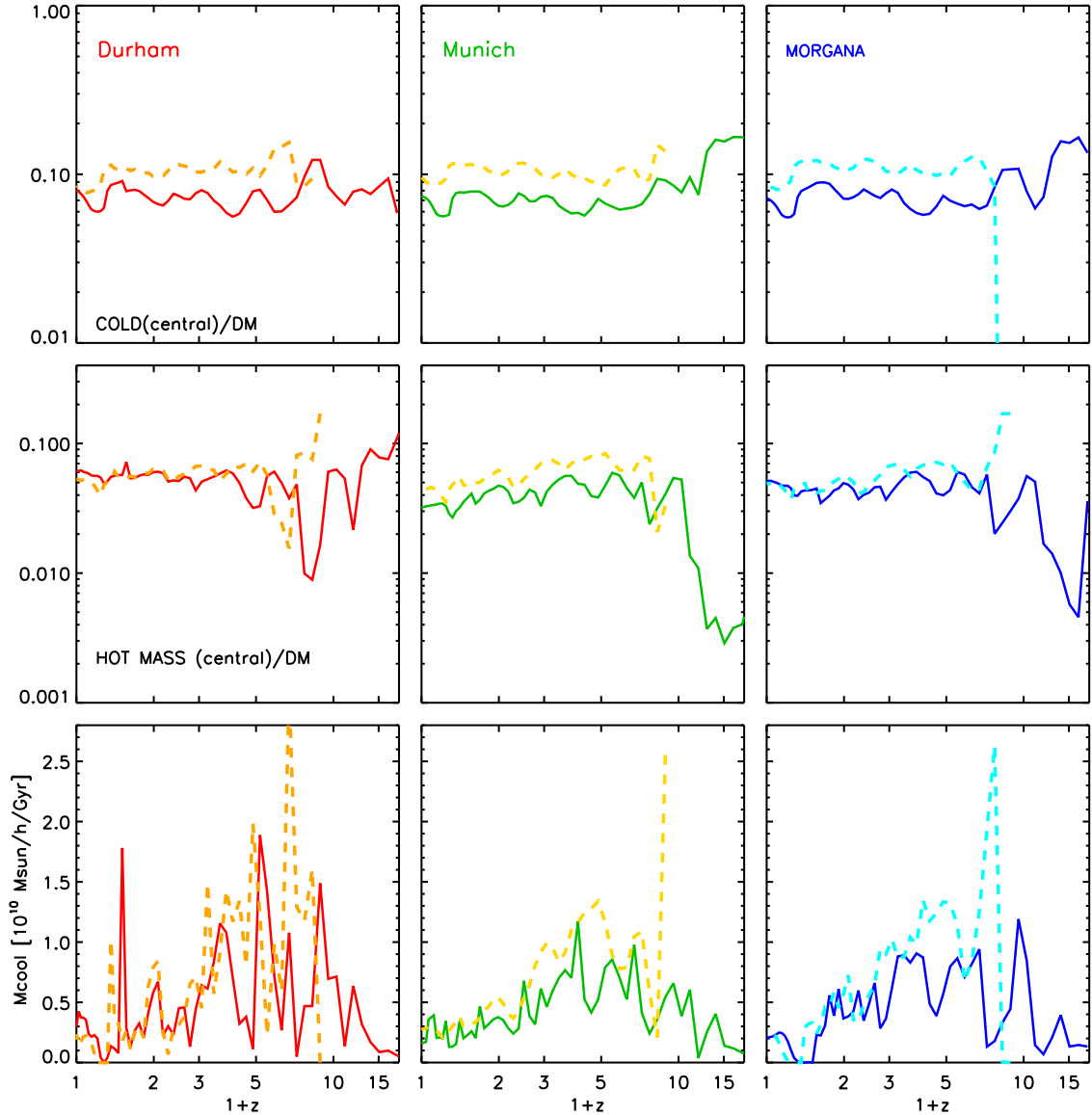}}
\caption{As in Fig.~\ref{fig:exmini1}. The mass accretion history of this halo
  is shown in the bottom panel of Fig.~\ref{fig:exminiDM}.}
\label{fig:exmini2}
\ec
\end{figure*}

Figs.~\ref{fig:exmini1} and \ref{fig:exmini2} show the evolution of the cold
and hot gas fractions (top and middle panels respectively) and the cooling
rates (bottom panels) for the two examples whose mass accretion histories are
shown in Fig.~\ref{fig:exminiDM}. In these figures, solid lines show the
predictions from the Durham, Munich and {\small MORGANA} models obtained using
the higher resolution trees from the MS-II, while dashed lines show
predictions from the lowest resolution simulations. For clarity, we have only
shown the cold gas fractions obtained when merger times are set to
infinity. Although the overall behaviour of cooling rates is similar in the two
resolution runs, particularly at low redshift, it is clear from these figures
that none of the models used in this study achieves a very good
convergence. The bottom panels of Figs.~\ref{fig:exmini1} and \ref{fig:exmini2}
show that at halo appearance in the lowest resolution simulation, all three
models generally predict a much larger cooling rate than obtained in the
highest resolution simulation. The cooling rates adjust rapidly at
approximately the levels predicted in the highest resolution simulation, but
are always somewhat larger than the higher resolution predictions. As a
consequence, cold gas fractions predicted using the lower resolution trees are
generally higher than obtained when using the higher resolution trees.

In its standard implementation, the Munich model employed in this study, runs
on subhalo-based merger trees, rather than on FOF-based trees like those that
we have used in previous sections. It is then interesting to ask how much
results discussed above are affected by the use of a different scheme for the
construction of the input merger trees. To address this question, we have run
the Munich model on all the subhalo trees corresponding to the FOF trees in
both the MW and {\small SCUBA} samples discussed in Section
\ref{sec:simstrees}. We remind the reader that the subhalo-based merger trees
for the MS and MS-II were constructed by \citet{Springel_etal_2005} and
\citet{Boylan-Kolchin_etal_2009} as summarized in Section \ref{sec:simstrees},
and that they are publicly available.

\begin{figure*}
\bc
\resizebox{15.5cm}{!}{\includegraphics[]{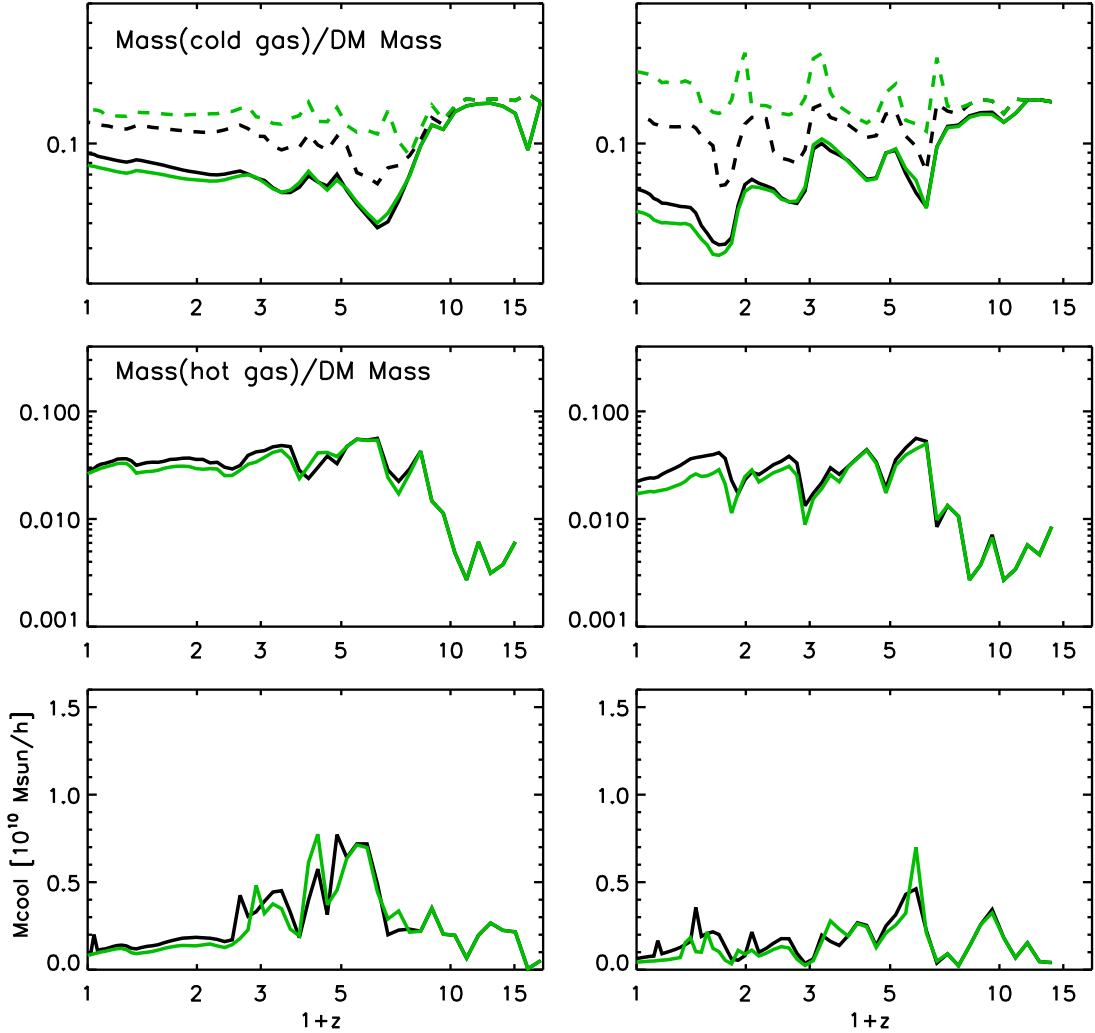}}
\caption{From top to bottom: cold gas fraction, hot gas fraction, and gas
  cooling rate for the same MW-like haloes shown in
  Fig.~\ref{fig:mwexamples}. Solid and dashed lines in the top panels show
  results for the $t_{mrg}=\infty$ and $t_{mrg}=0$ run, respectively. Black and
  green lines correspond to subhalo-based and FOF-based merger trees,
  respectively.}
\label{fig:ex_fofsubmw}
\ec
\end{figure*}

\begin{figure*}
\bc
\resizebox{15.5cm}{!}{\includegraphics[]{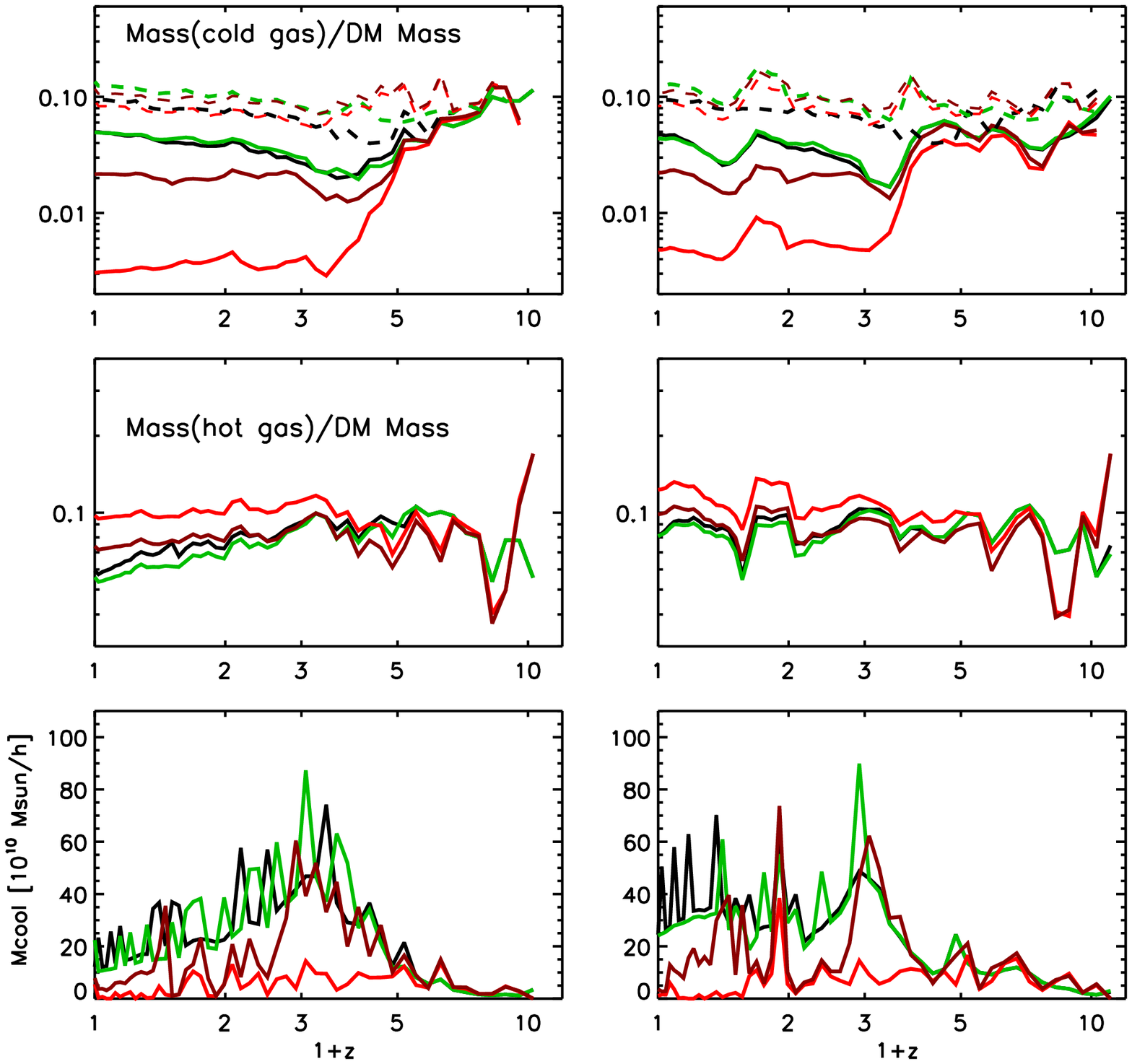}}
\caption{From top to bottom: cold gas fraction, hot gas fraction, and gas
  cooling rate for the same {\small SCUBA} haloes shown in
  Fig.~\ref{fig:scubaexamples}. Solid and dashed lines in the top panels show
  results for $t_{mrg}=\infty$ and $t_{mrg}=0$, respectively. Black and green
  lines correspond to subhalo based and FOF-based merger trees, while red lines
  correspond to the outputs from the Durham model, with darker lines showing
  results obtained assuming an isothermal profile for the hot gas.}
\label{fig:ex_fofsubscuba}
\ec
\end{figure*}

Figs.~\ref{fig:ex_fofsubmw} and \ref{fig:ex_fofsubscuba} show the evolution of
the cold gas fraction (top panels), hot gas fraction (middle panels) and
cooling rates (bottom panels) of the same MW and {\small SCUBA}-like examples
shown in Figs.~\ref{fig:mwexamples} and \ref{fig:scubaexamples}. In these
figures, green lines correspond to results from the FOF-based trees (and are
therefore equivalent to the results from the Munich model shown in
Figs.~\ref{fig:mwexamples} and \ref{fig:scubaexamples}), while black lines show
the corresponding results based on subhalo-based merger trees (red lines in
Fig.~\ref{fig:ex_fofsubscuba} will be discussed in Section \ref{sec:disc}
and can be ignored for now).

These figures show that predictions obtained using the FOF-based trees are in
quite nice agreement with those obtained using the subhalo-based trees. For
MW-like haloes, there is some tendency for the FOF-based trees to provide lower
cold and hot gas fractions with respect to results from subhalo-base trees. For
the {\small SCUBA} haloes, the cold gas fractions predicted from the two
schemes are very similar over all the redshift range where it is possible to
trace the main progenitor of the haloes under consideration. At $z<2$, the hot
gas fraction predicted using the FOF-based trees is only slightly lower than
results obtained using the standard (for the Munich model) subhalo scheme.

The bottom panels of Figs.~\ref{fig:ex_fofsubmw} and \ref{fig:ex_fofsubscuba}
show that the cooling rates obtained using the two different schemes are very
close, particularly at high redshift. Some small differences are, however,
visible.  These arise from different halo merger times due to the subhalo
scheme explicitly following haloes once they are accreted onto more massive
systems. We recall that in the standard Munich model, a residual merging time
(given by Eq.~\ref{eq:mergmunich}) is assigned to satellite galaxies only when
the dark matter substructures fall below the resolution limit of the
simulation. We have kept this assumption in the examples shown here so that the
$t_{mrg}=0$ case corresponds to galaxies merging at the time their parent
subhaloes are tidally stripped at the resolution limit of the simulation,
rather than instantaneously merging at the time their parent haloes become
proper substructures. That is why the two dashed lines shown in the top panels
of Figs.~\ref{fig:ex_fofsubmw} and \ref{fig:ex_fofsubscuba} follow a different
evolution. As we have explained in Section \ref{sec:models}, in this model the
amount of gas available for cooling is computed at the beginning of each
snapshot by assuming baryon conservation (see Eq.~\ref{eq:mhotcompute}). Since
then some cold gas is retained in distinct satellites for some time in the
subhalo scheme, this leads to a different amount of gas available for cooling,
and to the differences discussed above. It is interesting, however, that
despite the delay due to the identification of dark matter substructures, the
hot and cold gas fractions predicted by the two schemes, as well as the
predicted cooling rates, do not differ dramatically. The differences between
the two schemes become more important at lower redshift, but are in all cases
smaller than those obtained from alternative modelling schemes (see Sections
\ref{sec:mw} and \ref{sec:scuba}). This suggests that tidal stripping and
truncation are very efficient in reducing the high redshift substructures below
the resolution limit of the simulation, while dark matter subhaloes survive
longer as independent entities at lower redshift. This is expected, due to the
increase of dynamical times at later times (see also Fig.~4 in
\citealt{Weinmann_etal_2010}).

\section{Mergers}
\label{sec:mergtimes}

In this section, we compare the different implementations of galaxy mergers
adopted in the three models used in this study. To this aim, we have identified
all satellites that are present in each pair of models\footnote{Due to a
  different treatment of the rapid cooling regime, there are haloes that host a
  galaxy in one model and not in another. These `unpaired galaxies' are
  excluded in the analysis presented here.} and that are assigned merger times
that are lower than the Hubble time.  As a reminder, merger times in the Durham
model and in {\small MORGANA} are re-assigned after a new formation event, or
halo major merger. For the comparison discussed below, we have always
considered the last assignments in these two models.

\begin{figure*}
\bc
\resizebox{15.5cm}{!}{\includegraphics[angle=90]{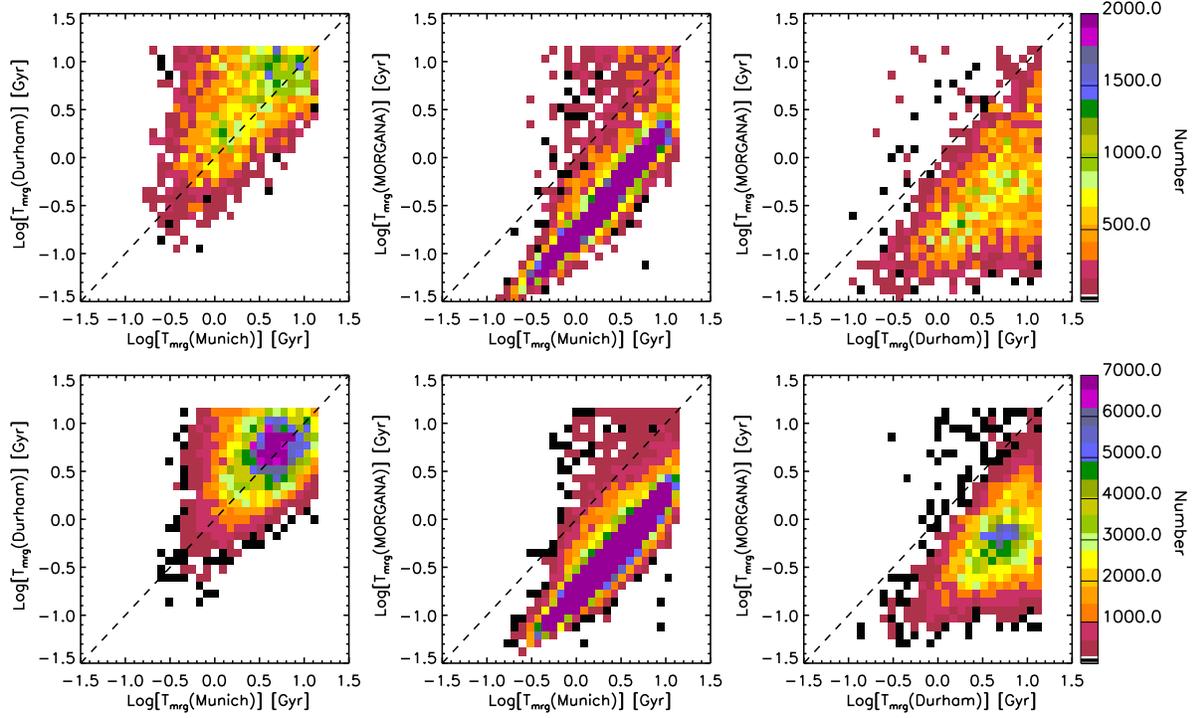}}
\caption{Distributions of merging times in the three models used in this
  study. Top panels are for MW haloes while bottom panels are for the SCUBA
  haloes. Maps have been computed using only satellite galaxies present in both
  models and with merger time smaller than the Hubble time.}
\label{fig:mergtimes}
\ec
\end{figure*}

Fig.~\ref{fig:mergtimes} compares the distributions of merging times in all
three models used in this study. Top panels are for satellites of the MW-like
haloes, while bottom panels are for {\small SCUBA}-like haloes. There is no
significant difference between the two samples, but a much larger number of
satellites for the {\small SCUBA} haloes. Fig.~\ref{fig:mergtimes} shows that a
fraction of model satellites distribute along the one-to-one relation when
comparing the merger times assigned in the Durham model with those obtained
from the Munich model (left panels). The scatter is, however, very large, with
a large number of satellites merging within relatively short times in the
Munich model and getting a much longer merger time in the Durham model, and
viceversa.  The correlation between the Munich model and {\small MORGANA} is
much tighter (middle panels in Fig.~\ref{fig:mergtimes}), but there is a clear
offset indicating that model satellites in {\small MORGANA} have merger times
that are systematically lower than the corresponding times from the Munich
model. Finally, when comparing the Durham model and {\small MORGANA} (right
panels), all satellites fall below the one-to-one relation with a quite large
scatter, and there is a large concentration of galaxies that merge within $\sim
5$~Gyr in the Durham model and within $\sim 0.6$~Gyr in {\small MORGANA}.

\begin{figure*}
\bc
\resizebox{15.5cm}{!}{\includegraphics[angle=90]{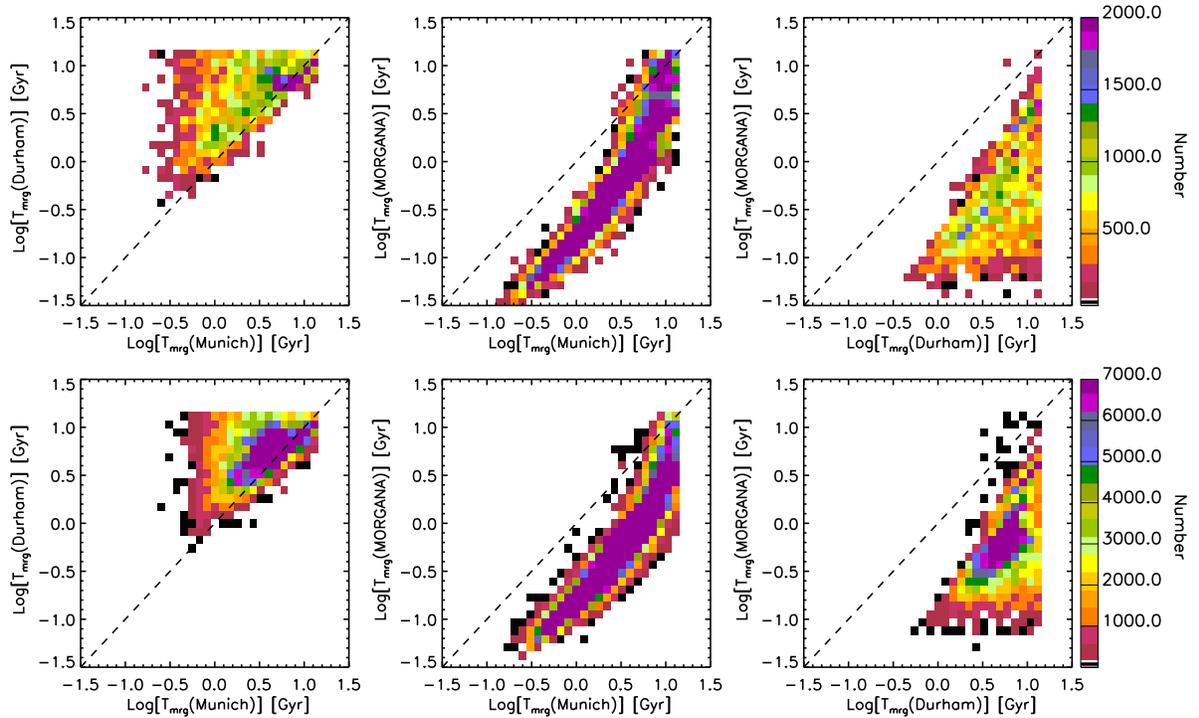}}
\caption{As in Fig.~\ref{fig:mergtimes}, but neglecting the orbital dependencies
  in the merger time calculation for the Durham and {\small MORGANA} models
  (see text for details).}
\label{fig:mergtimesnoorbit}
\ec
\end{figure*}

At least part of the scatter in Fig.~\ref{fig:mergtimes} is due to the random
extraction of orbital parameters in the Durham and {\small MORGANA} models. To
show a `cleaner' comparison, we have re-run both models neglecting the orbital
dependency. In the Durham model, this has been achieved by setting $\Theta_{\rm
  orbit} = 1$, while in {\small MORGANA} only circular orbits have been
considered, with orbital energy set to $0.5$ (see Section
\ref{sec:morgana}). Results are shown in Fig.~\ref{fig:mergtimesnoorbit}. As
expected, the scatter is reduced in all panels. A larger fraction of model
satellites distribute along the one-to-one relation in the left panels. There
is, however, still a significant fraction of satellites that get longer merger
times in the Durham model than in the Munich model. As discussed in Section
\ref{sec:moddiff}, the dynamical friction formulations adopted in the Munich
and Durham models differ by the assumptions made for the Coulomb logarithm, and
for a different numerical pre-factor. All differences visible in the left
panels of Fig.~\ref{fig:mergtimesnoorbit} are due to these different
assumptions, and to the re-assignment of merger times after each formation event
(see below).

There is still a strong correlation between the merger times assigned in
{\small MORGANA} and in the Munich model, with an offset towards lower merger
times in {\small MORGANA}. Neglecting the orbital dependency, the offset
reduces at long merger times. The correlation between {\small MORGANA} and the
Durham model is still quite bad, with a large fraction of satellites that merge
within $0.6$~Gyr in {\small MORGANA} and that are assigned merger times that
are about ten times longer in the Durham model.

\begin{figure}
\bc
\resizebox{8cm}{!}{\includegraphics[]{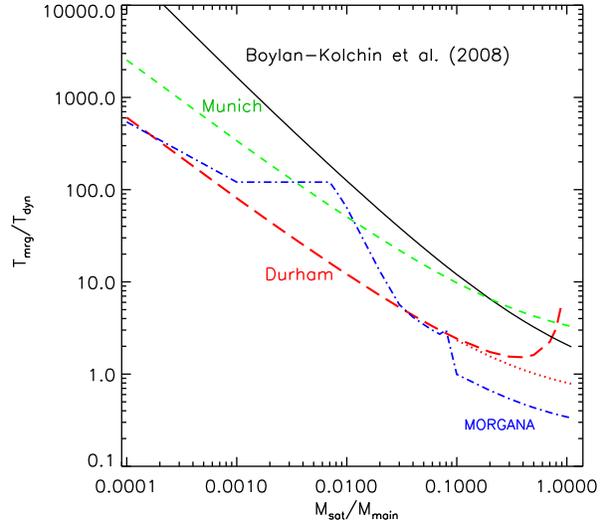}}
\caption{Merger times (in units of dynamical times) for different mass
  ratios. Dashed, long-dashed and dot-dashed lines correspond to the standard
  assumptions adopted by the Munich, Durham and {\small MORGANA} models
  respectively. The {dotted} line has been obtained from the long-dashed line
  by changing the assumption for the Coulomb logarithm to $\Lambda_{\rm
    df}=1+M_{\rm main}/M_{\rm sat}$.  The predictions from {\small MORGANA} do
  not depend significantly on halo concentration. The thick solid line
  corresponds to the fitting formula provided by
  \citet{Boylan-Kolchin_Ma_Quataert_2008}, with no orbital dependency and for
  circular orbits.}
\label{fig:compareDF}
\ec
\end{figure}

The results discussed above can be understood by comparing
Eqs. \ref{eq:mergmunich}, \ref{eq:mergdurham}, and \ref{eq:mergmorgana}. This
is done in Fig.~\ref{fig:compareDF}, which shows the merger times obtained
using these different formulations, for different mass ratios. In this figure,
we have neglected the orbital dependencies and have assumed that halo
concentration (this enters the equations adopted in {\small MORGANA}) varies as
a function of mass according to Eq.~4 in \citet{Neto_etal_2007}. The figure
shows clearly that the Munich and Durham implementations differ by a scaling
factor but for $M_{\rm sat}/M_{\rm main} \geq 0.1$, due to a different
assumption about the Coulomb logarithm: the dotted line in
Fig.~\ref{fig:compareDF} has been obtained from the long-dashed line by
changing the assumption for the Coulomb logarithm to $\Lambda_{\rm df}=1+M_{\rm
  main}/M_{\rm sat}$, as in the Munich model. The re-calculation of merger
times after each formation event or major merger apparently does not affect
significantly the expected disagreement. It should be noted that in
Figs.~\ref{fig:mergtimes} and \ref{fig:mergtimesnoorbit}, we have only used
satellite galaxies with merger times lower than the Hubble times. These figures
are therefore dominated by mergers with $M_{\rm sat}/M_{\rm main} \geq 0.1$. In
this regime, the {\small MORGANA} model is offset low with respect to the
Munich model. It also predicts systematically lower merger times with respect
to the Durham model but in this case the offset is not linear as a function of
the mass ratio. In Fig.~\ref{fig:compareDF}, we have also shown the fitting
formula provided by \citet{Boylan-Kolchin_Ma_Quataert_2008}, with no orbital
dependency and for circular orbits. As noted earlier, the adoption of a simple
fudge factor does not suffice to bring the formulation adopted in the Munich
model in agreement with the new fitting formula proposed by Boylan-Kolchin et
al.

\section{Summary and Discussion}
\label{sec:disc}

In this work we have compared results from three independently developed
semi-analytic models, and we have focused on alternative implementations of gas
cooling and galaxy mergers. Our model comparison is carried out using two large
samples of {\it identical} merger trees, which allows us to compare results on
a case-by-case basis and to focus on differences due to different model
assumptions and parametrizations. In particular, we have constructed two
FOF-based merger tree samples. One is a set of 100 haloes from the MS-II with
redshift zero dark matter halo masses similar to that of the Milky Way (MW-like
haloes), while the other consists of 100 haloes from the MS with a number
density similar to that measured for {\small SCUBA} galaxies at $z\sim 2$
({\small SCUBA}-like haloes). By using stripped-down versions of the models, we
are able to avoid possible degeneracies and complications due to a different
treatment of various physical processes and to concentrate on the influence of
specific assumptions and/or parametrizations. As explained above, we have
chosen to include only the processes of gas cooling and galaxy mergers. In the
following, we summarize briefly the results obtained and discuss them.

\subsection{Gas cooling}

Our results show that the different assumptions adopted about gas cooling lead
to very similar results on mass scales similar to those of our own Galaxy, and
to a generally good statistical agreement. Important differences arise,
however, on larger mass scales. 

Two of the models used in this study (the Munich and Durham models) assume
variations of the cooling model originally proposed by
\citet{White_and_Frenk_1991}. The other model used here ({\small MORGANA})
adopts a more sophisticated model and, on the basis of previous published
results \citep{Viola_etal_2008}, was expected to provide systematically higher
cooling rates for our {\small SCUBA}-like haloes. Contrary to this naive
expectation, however, results from the Munich model and from {\small MORGANA}
are very similar for this mass scale, while the Durham model gives
systematically lower cooling rates. These results appear to be in contradiction
with previously published tests, but the contradiction is only apparent as we
discuss below. 

As explained in detail in Section \ref{sec:models}, although the Munich and
Durham models are variations of the same cooling model, they differ in a number
of details, in particular for the assumption adopted for the hot gas
distribution. Red lines in Fig.~\ref{fig:ex_fofsubscuba} show results from the
standard Durham model (that assumes a $\beta$ profile for the hot gas) and from
a model that uses the same cooling implementation but assumes an isothermal
distribution for the hot gas (darker red lines), as in the Munich model. The
figure shows clearly that by changing this assumption, the predicted cooling
rates are larger, bringing model results closer to (although they are still
lower than) those obtained from the Munich model (green lines in
Fig.~\ref{fig:ex_fofsubscuba}). Some differences are still apparent, however:
the Munich model predicts systematically higher cooling rates than the Durham
model, particularly at high redshift, where cooling is much more efficient in
this model than in both of the Durham implementations considered here.  These
residual differences are due to a number of other different assumptions, in
particular for the calculation of the cooling radius (see Section
\ref{sec:moddiff}). Statistically, the differences between the models are
relatively {\it small}, which reflects a general agreement in the underlying
framework of these two models.

Perhaps more surprising is the similarity between the results obtained from the
Munich model and those from {\small MORGANA}, at both mass scales analysed in
this paper. As discussed earlier, \citet{Viola_etal_2008} have claimed that the
cooling model implemented in {\small MORGANA} predicts cooling rates that are
significantly larger than those obtained using the classical model by
\citet{White_and_Frenk_1991}. It should be noted, however, that their
implementation of the classical model did not assume any special treatment for
the `rapid cooling regime' (as is instead done in the original work by White
and Frenk and all subsequent variants of this model), and assumed that the hot
gas distribution is described by a polytropic equation of state with index
$\gamma_p = 1.15$, which is similar to the $\beta$ profile assumed in the
Durham model. In addition, results discussed in Viola et al. were obtained for
isolated static haloes and it is not trivial to generalize them to the case of
cosmological mass accretion histories, like those we have used here. Their
conclusions is, however, valid when one adopts a gas distribution similar to
that adopted in the Durham model used in this study (which is indeed similar to
that they assumed in their implementation of the classical model). Adopting a
steeper gas profile, as in the Munich model used here, changes results
significantly in some mass regimes, bringing them in very good agreement with
those from the {\small MORGANA} model.

It is important to realize that the differences highlighted above will have
important consequences on predictions from galaxy formation models.  For
example, the Munich and {\small MORGANA} models will need to assume a stronger
feedback than the Durham model, to counter-act excessive cooling at low
redshift in relatively large haloes. Although all models achieve this using
very similar schemes (the AGN feedback), the relative importance of these
additional physical processes will be different in these three
models. Predicting much lower cooling rates for massive haloes, the Durham
model will have difficulties in providing large numbers of galaxies with
elevated star formation at high redshift. Indeed, this model is able to
reproduce the observed galaxy number counts at $850\mu$m only by assuming a
top-heavy initial mass function (IMF) from the stars formed in bursts. This
works because a top-heavy IMF has a much larger recycled gas fraction, which
provides fuel for star formation. As illustrated above, {\small MORGANA}
predicts much larger cooling rates than the Durham model and is indeed able to
reproduce the number counts of submillimiter sources (but the brightest ones)
with a standard IMF \citep{Fontanot_etal_2007}. The Munich model used in this
study predicts even larger cooling rates on average, but its predictions for
the submillimiter number counts have not been explored yet.

\subsection{Galaxy mergers}

The three models used in this study adopt a different modelling for galaxy
mergers. The Munich and Durham model assume variations of the classical
dynamical friction formula. Results from these two models are somewhat
correlated, but there is a large scatter and a large number of galaxies get
significantly longer merger times in one model than in the other. As explained
in Section \ref{sec:mergtimes}, this is mainly due to the adoption of a
different numerical factor in front of the dynamical friction formula employed,
and to a different assumption about the Coulomb logarithm. {\small MORGANA}
uses formulae derived from numerical simulations and analytic models that take
into account dynamical friction, mass loss by tidal stripping, tidal disruption
of subhaloes, and tidal shocks \citep{Taffoni_etal_2003}. As shown above, over
the range of mass-ratios that provide merger times lower than the Hubble time,
the merger times obtained using these formulae are systematically lower (by a
factor $\sim 10 \, \tdyn$) than those computed from the Munich model. Of the
three models used in this study, the Munich model uses merger times that are
closer to the fitting formula recently proposed by
\citet{Boylan-Kolchin_Ma_Quataert_2008}, when neglecting any orbital
dependence.

The differences just discussed have important consequences for the stellar
assembly history of massive galaxies, and for the formation and evolution of
the brightest cluster galaxies and of the intra-group and intra-cluster light. 
{\small MORGANA} (and to some extent also the Durham model) will tend to
assemble massive central galaxies earlier than the Munich model. To keep these
galaxies red, these models will need to assume a somewhat stronger supernovae
feedback so as to make most of the mergers driving their late stellar assemble
{\it dry} (i.e. avoid triggering late bursts that would rejuvenate the stellar
population of these galaxies). A different balance between AGN cooling and tidal
stripping of stars will also be required in these models to keep model
predictions in agreement with observational results. These considerations are
of course valid in the case all models would use the same treatment of all
other physical processes at play. We remind, however, that as mentioned in
Section \ref{sec:intro}, these processes are usually treated in a different way
complicating the comparison between different models.

\subsection{Numerical resolution and merger tree scheme}

Taking advantage of the mini-MSII, run using the same initial conditions and
volume as for the MS-II but lower resolutions, we have analysed how the results
discussed above vary as a function of numerical resolution. The level of
agreement between the three models used in our study is not affected by
numerical resolution. None of the models used in this paper, however, achieves
a good numerical convergence, with all of them predicting moderately larger
cooling rates in lower resolution runs. 

On the other hand, results seem to be quite stable to alternative schemes for
the construction of dark matter merger trees. In particular, we have compared
results obtained using FOF-based trees with those obtained using subhalo-based
trees, which represent the standard input of the Munich model used in this
study. The small differences found by comparing results from these two schemes
can be ascribed to the possibility of tracking the accreted haloes until they
are stripped below the resolution limit of the simulation by tidal stripping
and truncation. Interestingly, our results show that these processes act on
relatively short time-scales, and that they are more efficient at higher
redshift.

\subsection{What Next?}
\label{sec:concl}

One question that this study does not address is: what is the {\it best} way to
model gas cooling and galaxy mergers? This is a very difficult question,
particularly for the cooling model. It is clear that one needs to understand
how the gas is distributed in dark matter haloes, and how this distribution is
affected by heating from supernovae and/or AGN feedback. Although
hydrodynamical simulations of galaxy formation are becoming increasingly
sophisticated, these physical processes still need to be included as `sub-grid'
physics, i.e. using prescriptions that are `semi-analytical' in nature. As a
consequence, published hydrodynamical simulations offer little indication of
appropriate modelling of the hot gas distribution and evolution.

Our results have pointed out that modelling of the rapid cooling regime differ
significantly in the implementations discussed in this paper. Substantial
numerical work has been focused recently on this mode of accretion, although it
was was discussed as early as \citet{Binney_1977},
\citet{Rees_and_Ostriker_1977}, and
\citet{White_and_Frenk_1991}. Interestingly, recent studies show that gas
accretion during the `quasi-static regime' in hydrodynamical simulations is
sensitive to different implementations of SPH \citep[][see also
  \citealt{Yoshida_et_al_2002}]{Keres_etal_2009}, while accretion rates in the
rapid cooling regime are quite robust. One possible additional concern in using
numerical results to inform semi-analytic models is the poor performance of SPH
codes in resolving and treating dynamical instabilities developing at sharp
interfaces in a multi-phase fluid \citep{Argetz_etal_2007}.

The situation is somewhat better for the modelling of galaxy mergers. Since the
merging process is predominantly driven by gravity, it can be studied using
controlled numerical experiments as done, for example, by
\citet{Boylan-Kolchin_Ma_Quataert_2008}. As noted earlier, however, recent work
has not yet converged on the dynamical friction formula appropriate for galaxy
formation models. Further work in this area is therefore needed.

Gas cooling and galaxy mergers are two basic ingredients of any galaxy
formation model that are relatively well understood. Also at this level,
however, different assumptions have to be made when implementing these
processes. These give rise to non-negligible differences that can have
important implications on the weight that needs to be given to additional
physical processes (e.g. AGN feedback, tidal stripping of stars, etc.). This
paper highlights specific areas where further work is needed in order to
improve our galaxy formation models, with the ultimate goal of improving our
understanding of the physical processes driving galaxy formation and evolution.

\section*{Acknowledgements}
GDL acknowledges financial support from the European Research Council under the
European Community's Seventh Framework Programme (FP7/2007-2013)/ERC grant
agreement n. 202781. AJB acknowledges the support of the Gordon \& Betty Moore
Foundation. GDL, AJB, and FF acknowledge the hospitality of the Kavli Institute
for Theoretical Physics of Santa Barbara, where the initial calculations that
led to this paper were carried out. The Millennium and Millennium-II Simulation
databases used in this paper and the web application providing online access to
them were constructed as part of the activities of the German Astrophysical
Virtual Observatory. We are grateful to Gerard Lemson for setting up an
internal database that greatly facilitated the exchange of data and information
needed to carry out this project. We are grateful to Carlton Baugh, Richard
Bower, Shaun Cole, Carlos Frenk and Cedric Lacey for making available the
Galform code for this project.

\bsp

\label{lastpage}

\bibliographystyle{mn2e}
\bibliography{sam_compare}

\end{document}